\documentclass[useAMS,usenatbib]{mn2e}

\usepackage{graphics}
\usepackage[dvips]{epsfig}
\usepackage{lscape}
\usepackage{multirow}
\usepackage{natbib}
\usepackage{aas_macros}
\usepackage{amsmath,amssymb}
\usepackage{epsfig}
\usepackage[table]{xcolor}
\def\grb{GRB\,070125}

\newcommand{\hi}{\mbox{H\,{\sc i}}}

\newcommand{\feii}{\mbox{Fe\,{\sc ii}}}

\newcommand{\siii}{\mbox{Si\,{\sc ii}}}

\newcommand{\nv}{\mbox{N\,{\sc v}}}
\newcommand{\niii}{\mbox{Ni\,{\sc ii}}}
\newcommand{\mgii}{\mbox{Mg\,{\sc ii}}}
\newcommand{\mgi}{\mbox{Mg\,{\sc i}}}
\newcommand{\cii}{\mbox{C\,{\sc ii}}}

\newcommand{\civ}{\mbox{C\,{\sc iv}}} 
\def\lya{Ly$\alpha$}

\title[\grb{} and the spectral-line poor absorbers]
{\grb{} and the environments of spectral-line poor afterglow absorbers\thanks{Based on target-of-opportunity observations collected in service mode under program ID 078.D-0416, P.I. Vreeswijk, with the FOcal Reducer/low dispersion Spectrograph 2 \citep[FORS2][]{Appenzeller98} installed at the Cassegrain focus of the Very Large Telescope (VLT), Unit 1, Antu, operated by the European Southern Observatory (ESO) on Cerro Paranal in Chile. For further information or questions on the content of the paper, please e-mail annalisa@raunvis.hi.is.}}

\author[De Cia et al.]
{A. De Cia$^{1}$, R. L. C. Starling$^{2}$, K. Wiersema$^{2}$, A. J. van der Horst$^{3}$, \and P. M. Vreeswijk$^{1}$, G. Bj\"{o}rnsson$^{1}$, A. de Ugarte Postigo$^{4}$, P. Jakobsson$^{1}$,\and A. J. Levan$^{5}$, E. Rol$^{6}$, S. Schulze$^{1}$ and N. R. Tanvir$^{2}$\\
$^{1}$ Centre for Astrophysics and Cosmology, Science Institute, University of Iceland, Dunhaga 5, IS-107 Reykjavik, Iceland\\ 
$^{2}$ Department of Physics and Astronomy, University of Leicester, University Road, Leicester LE1 7RH, UK \\
$^{3}$ Universities Space Research Association, NSSTC, 320 Sparkman Drive, Huntsville, AL 35805, USA\\  
$^{4}$ Dark Cosmology Centre, Niels Bohr Institute, University of Copenhagen, 2100 Copenhagen \O, Denmark\\
$^{5}$ Department of Physics, University of Warwick, Coventry, CV4 7AL, UK\\
$^{6}$ Astronomical Institute, University of Amsterdam, Science Park 904, 1098 XH Amsterdam, The Netherlands\\  
}

\begin{document}

\date{Accepted yyyy Month dd. Received yyyy Month dd; in original form yyyy Month dd}

\pagerange{\pageref{firstpage}--\pageref{lastpage}} \pubyear{2011}

\maketitle

\label{firstpage}

\begin{abstract}
\grb{} is among the most energetic bursts detected and the most extensively observed so far. Nevertheless, unresolved issues are still open in the literature on the physics of the afterglow and on the GRB environment. In particular, \grb{} was claimed to have exploded in a galactic halo environment, based on the uniqueness of the optical spectrum and the non-detection of an underlying host galaxy. In this work we collect all publicly available data and address these issues by modelling the NIR-to-X-ray spectral energy distribution (SED) and studying the high signal-to-noise VLT/FORS afterglow spectrum in comparison with a larger sample of GRB absorbers. The SED reveals a synchrotron cooling break in the UV, low equivalent hydrogen column density and little reddening caused by a LMC- or SMC-type extinction curve. From the weak \mgii{} absorption at $z=1.5477$ in the spectrum, we derived $\log N$(\mgii{}$)=12.96^{+0.13}_{-0.18}$ and upper limits on the ionic column density of several metals. These suggest that the GRB absorber is most likely a Lyman limit system with a $0.03\,Z_\odot <Z<1.3\,Z_\odot$ metallicity. The comparison with other GRB absorbers places \grb{} at the low end of the absorption line equivalent width distribution, confirming that weak spectral features and spectral-line poor absorbers are not so uncommon in afterglow spectra. Moreover, we show that the effect of photo-ionization on the gas surrounding the GRB, combined with a low $N$(\hi{}) along a short segment of the line of sight within the host galaxy, can explain the lack of spectral features in \grb{}. Finally, the non-detection of an underlying galaxy is consistent with a faint GRB host galaxy, well within the GRB host brightness distribution. Thus, the possibility that GRB 070125 is simply located in the outskirts of a gas-rich, massive star-forming region inside its small and faint host galaxy seems more likely than a gas-poor, halo environment origin.
\end{abstract}


\section{Introduction}

The progenitors of long gamma ray bursts (GRBs) are believed to be rapidly rotating massive stars \citep[for a review, see][]{Woosley93,MacFadyen99}. The observations of some SNe associated with GRBs have corroborated this hypothesis \citep[][and references therein]{Woosley06,Hjorth11}, implying that long GRBs are located in star-forming regions. Within the fireball model \citep[see][for a review]{Piran04,Meszaros06} the GRB afterglow is produced by the interaction of an expanding blast wave with the surrounding medium. The external shocks are responsible for the synchrotron radiation, from X-rays to optical and down to radio frequencies as the blast wave decelerates. The spectral energy distribution (SED) and its evolution is described by the fireball model \citep{Sari98,Granot02}. Modelling the SED can simultaneously provide information on the intrinsic spectral shape of the GRB, i.e. the fireball physical parameters, and on the GRB local and host galaxy environment. In the latter case, this includes the circumburst density or the line of sight dust extinction, depending on whether the broadband blast wave, down to radio frequencies, is modelled \citep[e.g.,][]{Panaitescu02,Bjornsson04,Postigo05,Vanderhorst05,Johannesson06,Rol07} or the optical-to-X-ray SED \citep{Starling07,Starling08,Schady10, Zafar11}. The interstellar medium of the GRB host galaxy can also be investigated through absorption line spectroscopy of the afterglow, providing the chemical composition of different ions in the absorbing gas and possibly its metallicity \citep[e.g.,][]{Vreeswijk04,Fynbo06,Savaglio06,Prochaska07,Ledoux09}. 

\grb{} is among the most energetic bursts detected so far, both in the prompt high energy release \citep[with an isotropic emitted energy $E_{\rm iso}\sim10^{54}$ erg:][]{Bellm08} and its afterglow \citep[with a collimation-corrected blast wave kinetic energy of more than $10^{52}$ erg:][]{Chandra08}. The optical afterglow is amongst the brightest at around one day after the burst, when redshift-corrected \citep[$R_c\sim17$ shifted at $z=1$:][]{Updike08,Kann10}. The brightness of the event allowed a global monitoring campaign of the afterglow resulting in one of the largest datasets collected so far, covering radio, mm, NIR, optical, UV and X-rays \citep{Cenko08,Chandra08,Dai08,Updike08,Kann10}. Despite the observational effort, there is still much debate over both the jet properties and the type of environment in which \grb{} was situated.

Based on the absence of strong spectral features and of a bright underlying host, \grb{} was reported to be the first long GRB to have exploded in a galactic halo environment. This would pose a challenge to modelling of progenitor objects. In particular, the GRB site was suggested to be in a compact stellar cluster that resulting from the interaction of two blue faint galaxies at a projected distance of $\approx 27$ and $\approx 46$ kpc\footnote{Derived from the $3.2\arcsec$ and $5.5\arcsec$ angular distance between the afterglow and the two putative host galaxies.} \citep{Cenko08}. However, a very limited sample of GRB afterglows was used as reference to argue the peculiarity of the spectrum (few spectral features, weak \mgii{} absorption and low inferred $N$(\hi{})). By contrast, from a partial broadband SED analysis, \citet{Chandra08} found a very high circumburst density ($n\approx50$ cm$^{-3}$, from the kinetic energy and $n\approx40$ cm$^{-3}$, from the broadband modelling of the SED). However, in their data it was not possible to distinguish between a constant or a wind-like density profile nor to firmly constrain the putative (chromatic) jet break and the inverse Compton influence on it. On the other hand, \citet{Dai08} claimed a ``textbook'' achromatic jet break at 5.8 days after the burst. Another open issue is the presence or absence of a synchrotron cooling break in the SED between the optical and X-rays and the agreement or not of the data with the fireball closure, as discussed in \citep{Updike08}.

In this work, we collect the multitude of observations spread through the literature and analyse them together in order to obtain a coherent picture of the GRB environment. In particular, we investigate the location of a synchrotron cooling break and the environment dust and metal content by studying the NIR-to-X-ray SED at different epochs. Furthermore, we analyse the very high signal-to-noise (S/N) VLT/FORS spectrum of the afterglow and discuss the possible implications for the absorber environment. We then compare the spectral features to the large sample of GRB afterglows with redshift collected in \citet{Fynbo09} and two sub-samples of GRBs potentially similar to \grb{}. The observations and data analysis are presented in Sec. \ref{sec obs} and Sec. \ref{sec analysis} respectively, the SED modelling in Sec. \ref{sec mod} and the discussion in Sec. \ref{sec discussion}. Our conclusions are summarized in Sec. \ref{sec conclusions}.  

Throughout the paper we use the convention $F_\nu \left(t\right) \propto t^{-\alpha}\nu^{-\beta}$ for the flux density  of the afterglow, where $\alpha$ is the temporal slope and $\beta$ is the spectral slope. We refer to the Solar abundances measured by \citet{Asplund09} and adopt cm$^{-2}$, as the linear unit of column densities, $N$. Hereafter, we assume a standard $\Lambda$CDM cosmology with $H_0$ = 70.4 km s$^{-1}$ Mpc$^{-1}$, $\Omega_M$ = 0.27 and $\Omega_{\Lambda}$ = 0.73 \citep{Jarosik11}.

\section{Observations and data reduction}
\label{sec obs}

\grb{} was localized by the Interplanetary Network (\textit{Mars Odyssey}, \textit{Suzaku}, \textit{INTEGRAL} and \textit{RHESSI}) at $t_0=$07:20:45 UT \citep{Hurley07}, and subsequently observed by \textit{Swift}. The high energy properties of the GRB prompt emission are discussed in \citet{Bellm08}. The optical afterglow was first localized by \citet{Cenko07a} at the position R.A. $=07^\mathrm{h}51^\mathrm{m}17.75^\mathrm{s}$ and Dec. $=+31^\circ09\arcmin04.2\arcsec$ (J2000) and the host galaxy redshift, $z=1.547$, was determined from a \mgii{} doublet absorption in a Gemini-North/GMOS afterglow spectrum \citep{Fox07}. Thanks to the brightness of the afterglow, a multitude of observations in the optical and NIR could be secured with 20 telescopes \citep[e.g.,][]{Chandra08,Updike08,Kann10}.

\begin{figure*}
\includegraphics[width=180mm,angle=0]{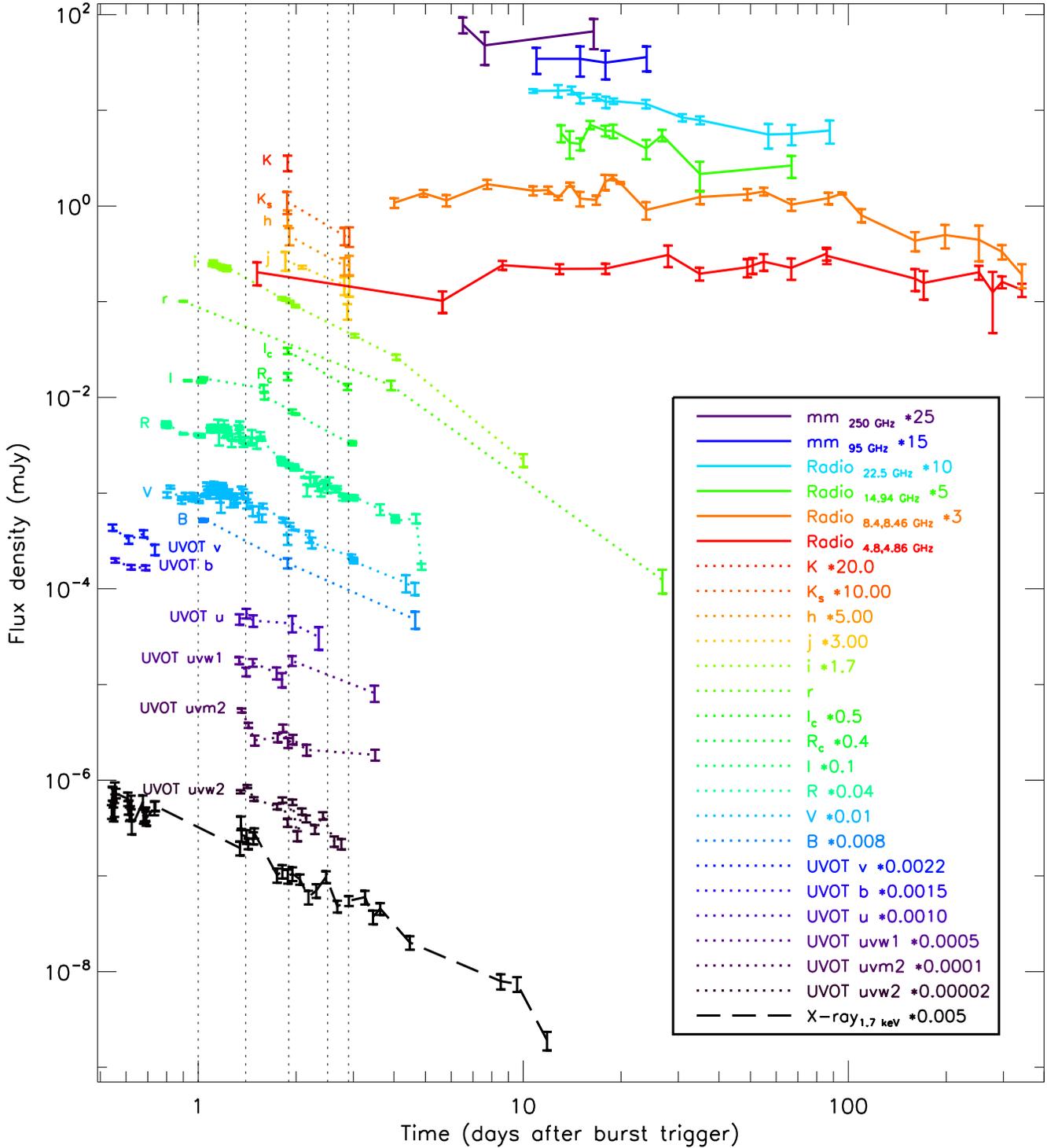}
\caption{The radio to X-ray afterglow light curves, scaled with the factors reported in the legend for a clearer display. The dashed vertical lines mark the SED times.}
\label{fig:alllc}
\end{figure*}

\subsection{\textit{Swift} XRT}

The \textit{Swift} X-ray telescope \citep[XRT,][]{Burrows05} began observing the \grb{} field 12.97 hrs after the trigger and continued to monitor the decaying afterglow in photon counting (PC) mode for 15 orbits or 173 ks of exposure time. The last observations ended on 2007 February 12 at 23:50 UT, 18.69 days after the burst trigger. We obtained the XRT event files from the UK {\it Swift} Science Data Centre archive\footnote{\texttt{http://www.swift.ac.uk/swift\_portal}} and processed them using the standard {\it Swift} XRT pipeline software version 0.12.3 and calibration database CALDB-11.

We extracted X-ray spectra and response files for each orbit individually as well as a late-time combined spectrum comprising orbit 3 onwards. Spectra were grouped such that a minimum of 20 counts fell in each bin and spectral analysis was performed with the\begin{small} XSPEC\end{small} spectral fitting package \citep{Arnaud96}, version 12.6.0.

We extracted a 0.3--10 keV light curve and transformed the count rate to flux using 3 conversion factors derived from spectra of orbit 1, orbit 2 and orbit 3 onwards of 1 ct s$^{-1} = 4.44, 4.09, 3.67 \times 10^{-11}$ erg cm$^{-2}$ s$^{-1}$, respectively. These three intervals were selected in order to apply the proper conversion factor to each regime, from the brighter to the late fainter epochs.  We also collected the dynamically binned XRT flux light curve from the \textit{Swift} repository \citep{Evans07}, which was converted from count rate to flux with the average conversion factor, 1 ct s$^{-1}$ $=3.8 \times10^{-11}$ergs cm$^{-2}$ s$^{-1}$. We then converted the binned and unbinned light curves into flux density $F_\nu$ at 1.73 keV, the logarithmic average of the XRT band (0.3--10 keV), assuming a spectral slope $\beta=1.1$, as found in Sect. \ref{sec x-rays}. The binned XRT light curve is shown in Fig. \ref{fig:alllc}.

\subsection{\textit{Swift} UVOT}

The afterglow of \grb{} was detected with the \textit{Swift} ultraviolet/optical telescope \citep[UVOT,][]{Roming05} in all six lenticular filters spanning $v$ to $uvw2$ (covering a range between 546--193 nm). These observations were coincident with those of the {\it Swift} XRT during the period 12.97 hrs to 18.69 days after the trigger. The optical/UV transient was detected out to 3.5 days after the trigger. We inspected the UVOT sky images and selected those of sufficient quality to perform photometry to an accuracy of 0.4 magnitudes or better. We performed photometry on the images using the {\sc uvotsource} task included in  \begin{small}HEASOFT\end{small} v6.7 and calibrations provided by \citet{Poole08}. We used a 5$\arcsec$ radius source extraction region, and a nearby 20$\arcsec$ radius source-free background region. The UVOT photometry is reported in the appendix

\subsection{VLT/FORS2}

Starting at 04:25 UT on 2007 January 26  (21.1 hours post-burst), a series of spectra were obtained with VLT/FORS2 in long-slit spectroscopy mode with a 1\farcs0 wide slit. The sequence of grisms used was 300V, 600z+OG590, 1400V and 1200R+GG435. This allowed us to both cover a larger wavelength window with the lower resolution grism (300V) and obtain mid-resolution spectroscopy for different regions of the spectrum. Cosmic ray removal was performed on each individual spectrum using the Laplacian Cosmic Ray Identification algorithm of \citet{vanDokkum01}. The seeing remained relatively stable during the observations, between 0\farcs6 and 0\farcs9, yielding the spectral resolutions and excellent S/N, reported in Table \ref{tab_log}.

\begin{table*}
\centering
\caption{VLT/FORS2 observation log on date 2007 January 26. $^a$ Mid-exposure time after the burst trigger (07:20:45 UT).} 
\begin{tabular}{ l | c c c c c c c c}
\hline \hline
\rule[-0.2cm]{0mm}{0.6cm}
Start time     & $\Delta\,t^a$ & Exp. time &  Grism  & Coverage & Resolution  & Airmass & Seeing & S/N /pixel\\
\rule[-0.2cm]{0mm}{0.6cm}
UT (hh:mm:ss)    &   (hrs)            &   (min)      &        &$\lambda$ (\AA{})& $FWHM$ (\AA{}) && $\arcsec$& (mean) \\  
\hline
04:25:24& 21.20 & 15 & 300V & 3190--9640 & 10.7 &   1.8 &  0.80&54 \\
04:53:31& 21.80 & 30 & 600z & 7460--10740 &5.3 &   1.8 & 0.72&39\\
05:25:14& 22.32 & 30 & 1400V & 4630--5930 & 2.0 &   1.9 & 0.87&39 \\
05:56:40& 22.85 & 30 & 1200R & 5870--7370 &  2.2 &   2.1 & 0.62&43\\

\hline \hline
\end{tabular}

\label{tab_log}
\end{table*}

\subsection{Literature data collection}

We gathered all the data on \grb{} available in the literature from \citet{Chandra08}, \citet{Updike08} and \citet{Kann10}, covering radio to optical frequencies. We exclude the observations from GRB Coordinates Network (GCNs) circulars reported in \citet{Chandra08}, with the exception of three NIR datapoints\footnote{The three GCN measurements do not strongly influence the NIR light curves, since they are very close to the NIR photometry reported by \citet{Updike08} at 2.9 after the burst.} ($K_s, J, H$), given the limited NIR coverage. In addition, we include the \textit{Swift} UVOT and XRT observations that we analysed. In all, 365 NIR to UV photometric data points were collected, for a total of 19 filters, and 77 radio and mm data points. We converted the magnitudes into flux density using the zero-points available for each specific filter. All the magnitudes were corrected for Galactic extinction ($E(B-V) = 0.052$ mag), as listed by \citet{Schlegel98}, and for transmission through the Ly$\alpha$ forest at redshift 1.5477 in each optical and UV band \citep[e.g.,][]{Madau95} adopting the spectral slope, $\beta$=0.58, as derived from the optical SED \citep{Updike08}. The corresponding transmissions are $tr_U=0.996$, $tr_{uvw1} = 0.848$, $tr_{uvm2} = 0.633$ and $tr_{uvw2} = 0.539$. The light curves for each filter are shown in Fig. \ref{fig:alllc}. The dataset that we used is rather inhomogeneous, as it is collected from different sources in the literature. Despite the uncertainty on the single data points, the interpolation of the flux densities at a given epoch are weighted by these errors, giving more importance to the more precise measurements. Thus, the uncertainties due to the inhomogeneous sample are limited, given the large dataset, and cannot significantly influence our results.

\section{Spectral analysis}
\label{sec analysis}

\subsection{X-ray spectrum}
\label{sec x-rays}

We first independently modelled the spectrum of each orbit with a simple power law, absorbed by gas at redshift $z=1.5477$, as derived from the spectral analysis (see Sec. \ref{sec opt spec an}), finding no evidence for spectral variability in the XRT data alone within this model. The Galactic absorption was fixed at 4.3 $\times$ 10$^{20}$ cm$^{-2}$ \citep{Kalberla05}. We then extracted the time-averaged spectrum over the whole XRT observation. Modelling the overall spectrum we derived a power-law spectral slope $\beta=1.06^{+0.09}_{-0.13}$,  a soft X-ray absorption at the host galaxy redshift, $N$(H) $=15.5^{+13.9}_{-14.9} \times 10^{20}$ cm$^{-2}$ and $\chi^2_{\nu}= 0.7$ for 40 degrees of freedom (dof). The quoted errors correspond to 90\% confidence. The modelled absorbed flux (0.3--10 keV) is $F=1.83\times10^{-13}$ ergs cm$^{-2}$ s$^{-1}$. The uncertainty on the host $N$(H) is sufficiently large that it would be consistent with only Galactic foreground absorption. Given that the Galactic $N$(H) is typically $10^{20}\lesssim N(\mbox{H})\lesssim 10^{21}$ cm$^{-2}$ (with uncertainties of about $\sim10$\% \citep{Dickey90,Kalberla05}, it is often not possible to tightly constrain a low GRB host galaxy $N$(H).

\subsection{Optical spectrum}
\label{sec opt spec an}

\begin{figure}
\includegraphics[width=88mm,angle=0]{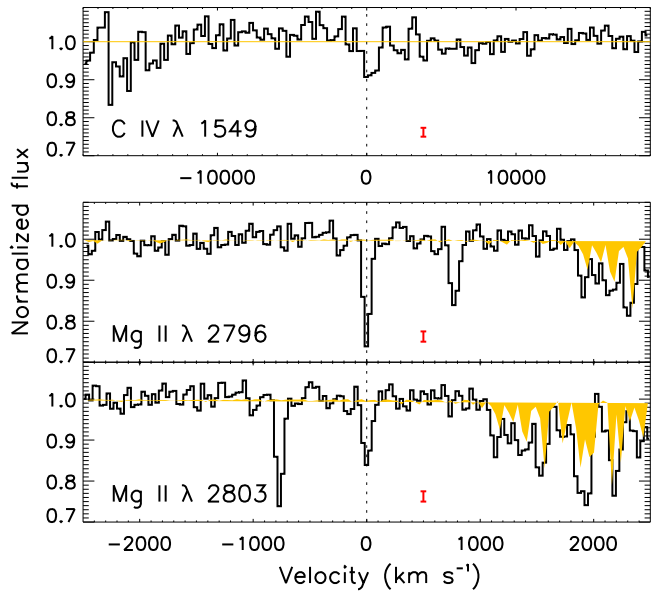}
\caption{Absorption lines detected in the FORS spectra at host galaxy redshift $z=1.5477$. Note that the \civ{} spectrum, observed with the lowest resolution, spans a larger velocity range (top panel). The error bars show the mean error spectrum, averaged within $\pm2000$ km s$^{-1}$ from the line position. The telluric features are highlighted.}
\label{fig det lines}
\end{figure}

The absorption lines detected in the FORS spectra are listed in Table \ref{tab_lines}, along with the equivalent widths ($EW$s). The absorption lines associated with the $z=1.5477$ system are shown in Fig. \ref{fig det lines}. The apertures used to measure the $EW$s are derived from the resolution in each grism, i.e. twice the full width at half maximum of the arc lines ($2\times FWHM_{\textrm{arc}}$), justified by the narrowness of both the \mgii{} and \civ{} lines. These lines are unresolved, the width of the \mgii{} $\lambda$ 2803 \AA{} line being $FWHM_{2803}\sim70$ km s$^{-1}$.

\begin{figure}
\includegraphics[width=8.5cm]{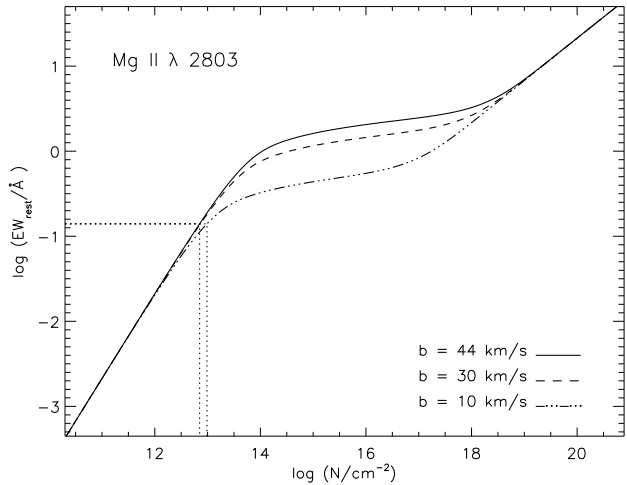}
\caption{Curve of growth for \mgii{} $\lambda$ 2803 for different $b$ values. The dotted lines correspond to the measured $EW_{\textrm{rest}}$. Including the $1\sigma$ errors on the $EW$, we derive log\,$N$(\mgii{}) $=12.96^{+ 0.13}_{-0.18}$.}
\label{fig cog}
\end{figure}

The low spectral resolution does not allow us to directly derive the column density from the detected $EW$ in the optically thin limit, even for the weak $EW$s we measure, since the lines may be saturated. In particular, this could affect the \mgii{}, despite the $EW_{2796}/EW_{2803}\sim 2$ ratio of the \mgii{} $\lambda$ 2796 and \mgii{} $\lambda$ 2803 \AA{} suggests little or no saturation. A proper curve of growth \citep[CoG, e.g.,][]{Mihalas78} analysis would require many transitions to constrain the $b$ value (Doppler parameter, $b=\sqrt{2} \times \sigma_{\textrm{Gaussian}}$). However, we can benefit from the excellent S/N and the narrowness of the transitions to constrain the column density with the CoG, despite the limited number of detections. We studied the CoG evolution with the $b$ value of the \mgii{} $\lambda\lambda$ 2796, 2803 transitions. The rest-frame $EW$ of the absorption lines fall between the linear and the saturated part of the CoG, i.e., the \mgii{} column density has a slight dependency on the $b$ value. We can limit the plausible range to $10<b<44$ km s$^{-1}$, as the line broadening must be narrower than the observed $FWHM$ $\sim70$ km s$^{-1}$ ($FWHM=2\sqrt{2 \ln 2}\,\, b/\sqrt{2}=1.665\,b$). The lower limit is selected from the lowest velocity width of low-ionization line profiles observed in UVES QSO absorbers\footnote{The lower observed value is $\Delta v=17$ km s$^{-1}$, where $\Delta v$ is the velocity width that covers 95\% of the complex line profile. If the line profile is a single Gaussian then $\Delta v=1.96\,\sigma_{\textrm{Gaussian}}=1.39\,b$, corresponding to a minimum observed $b=12$ km s$^{-1}$. This represent only an estimate of the width of the observed line profile, despite the single components of the complex profile not being resolved with low resolution spectroscopy. The UVES spectrum of GRB\,050730 showed single component line profiles with $b=10$ km s$^{-1}$. Of course, the possibility that the line profile is a single component with a very low $b$ can never be totally excluded. This would result in higher saturation of such component, leading to a higher value of $N$(\mgii{}). A higher $N$(\mgii{}) would only strengthen the conclusions of this paper. The \mgii{} column density derived in the optically thin limit is $\log N$(\mgii{})~$=12.88^{+0.06}_{-0.07}$.} \citep{Ledoux06b}. The $\lambda$ 2803 absorption line gives the tightest constraint, $\log N$(\mgii{}$)=12.96^{+ 0.13}_{-0.18}$ ($1\sigma$ errors), derived considering the median column density within the $b$ and $1\sigma$ $EW$ interval. The \mgii{} $\lambda$ 2803 CoG is shown in Fig. \ref{fig cog}. So far we used $EW_{2803}$ and the limits on $b$ to derive the \mgii{} column density. On the other hand, the \mgii{} column density derived from the $\lambda$ 2803 allows us to use the $\lambda$ 2796 CoG to further constrain the $b$ value to be $30< b<44$ km s$^{-1}$. With these limits on $b$ we cannot derive the \civ{} column density, because the $\lambda$ 1549 CoG shows that the absorption line is saturated. We list the $3\sigma$ upper limits on the non-detected lines in Table \ref{tab_ul}, measuring the noise level at each line position, again with $2\times FWHM_{\textrm{arc}}$ apertures. The excellent S/N allows us to put strict constraints on the column densities of some ions, such as \feii{}.

\begin{table}
\centering
\caption{Detected absorption lines in the VLT/FORS spectra. $^a$~1$\sigma$ errors. $^b$ Significance}
\begin{tabular}{ l c c c c c}
\hline \hline
\rule[-0.4cm]{0mm}{1.cm} 
Ion   & $\lambda_{\textrm{rest}}$ (\AA{})& $\lambda_{\textrm{obs}}$ (\AA{})& $z$ & $EW_{\textrm{rest}}^a$ (\AA{}) & $\sigma^b$\\
\hline
\multicolumn {6} { c } {300V} \\
\hline
\rule[-0.25cm]{0mm}{0.6cm} 
\mbox{C\,{\sc iv}}   &   1549   &  3949.55 & 1.5497 & $0.77\pm0.09$ &  8.5\\
\hline
\multicolumn{6}{c}{1200R} \\
\hline
\rule[-0.0cm]{0mm}{0.4cm}
\mbox{Mg\,{\sc ii}}   &   2796.35 & 7124.30 & 1.5477 &  $0.29\pm0.04$ & 7.2\\
\rule[-0.2cm]{0mm}{0.6cm}
\mbox{Mg\,{\sc ii}}   &   2803.53  &  7142.74 & 1.5478 & $0.14\pm0.02$ & 7.0\\
\hline
\multicolumn{6}{c}{1400V}\\
\hline
\rule[-0.0cm]{0mm}{0.4cm}
\mbox{Na\,{\sc i}} &  5891.58 & 5891.58 & 0.000 &0.68$\pm$0.08 & 8.5\\
\rule[-0.2cm]{0mm}{0.6cm}
\mbox{Na\,{\sc i}}  &  5897.56 & 5897.56 & 0.000 & 0.41$\pm$0.08 & 5.1\\

\hline \hline
\end{tabular}
\label{tab_lines}
\end{table}

\begin{table}
\centering
\caption{3$\sigma$ line detection limits with corresponding column densities computed in the optically thin limit.}
\begin{tabular}{ l c c c r}
\hline \hline
\rule[-0.4cm]{0mm}{1.cm} 
Ion   & $\lambda_{\textrm{rest}}$ (\AA{})& $EW_{\textrm{rest}}$ (\AA{}) & log N & Grism \\
\hline

\rule[-0.0cm]{0mm}{0.4cm}
\mbox{C\,{\sc ii}} & 1334.53   &  $<2.85$ & $<15.15$ & 300V \\
\mbox{Si\,{\sc iv}} & 1393.75  &  $<1.91$ & $<14.32$ & 300V \\
\mbox{Si\,{\sc ii}} & 1526.71   &  $<0.44$ & $<14.23$ & 300V \\
\mbox{Al\,{\sc ii}} & 1670.79   &  $<0.29$ & $<12.80$ & 300V \\
\mbox{Ni\,{\sc ii}} & 1741.55   &  $<0.27$ & $<14.37$ & 300V \\
\mbox{S\,{\sc i}} & 1807.31      &  $<0.25$ & $<13.89$ & 300V \\
\mbox{Al\,{\sc iii}} & 1854.72   &  $<0.09$ & $<12.76$ & 1400V \\
\mbox{Ti\,{\sc ii}} & 1910.93    &  $<0.07$ & $<13.36 $& 1400V \\
\mbox{Zn\,{\sc ii}} & 2026.14   &  $<0.06$ & $<12.55$ & 1400V \\
\mbox{Cr\,{\sc ii}} & 2056.25   &  $<0.07$ & $<13.23$ & 1400V \\
\mbox{Fe\,{\sc ii}} & 2382.76   &  $<0.07$ & $<12.61$ & 1200R \\
   \rule[-0.2cm]{0mm}{0.4cm}
\mbox{Fe\,{\sc ii}}* & 2396.35  &  $<$0.07 & $<$12.64 & 1200R\\

\hline\hline
\end{tabular}
\label{tab_ul}
\end{table}

We identified Galactic ($z=0$) absorption of the \mbox{Na\,{\sc i}} $\lambda\lambda$ 5891,5897 doublet in both 1200R and 1400V FORS spectra and we report the $EW$ in Table \ref{tab_lines}. The Galactic absorption is shown in Fig. \ref{fig Gal}. The Galactic reddening in the direction of the target measured by \citet{Schlegel98} is $E(B-V) = 0.052\pm0.008$ mag. For comparison, we derive the reddening from the Galactic \mbox{Na\,{\sc i}} - $E(B-V)$ correlation given by \citet{Munari97}, finding a higher reddening value, $E(B-V)\sim0.20$ mag, but consistent within the reddening uncertainty in their correlation (0.15 mag).

\section{Afterglow modelling: the NIR-to-X-ray SED}
\label{sec mod}

In order to investigate any temporal evolution of the spectral energy distribution (SED) from the NIR to X-ray frequencies, we evaluated the SED at five epochs after the burst trigger. The SED times 1.0, 1.4, 1.9, 2.5, 2.9 days were selected in order to maximize the NIR to UV data coverage for each epoch and minimize extrapolations, as shown in Fig. \ref{fig:alllc}. For each SED epoch, we combine the available NIR to UV photometric data points and the ones derived by interpolating or extrapolating the light curves where necessary, for each filter. The flux interpolations at particular times were performed locally, i.e., using up to three adjacent data points on each side. The X-ray spectra at the five epochs were produced by scaling the overall X-ray spectrum to the actual count-rate at each time of interest. This choice allowed to preserve the best S/N, justified by the lack of spectral evolution measured for the X-ray data alone (see Sec. \ref{sec x-rays}).\\ 

The SED was created and fitted in count space using the \begin{small}ISIS\end{small} spectral fitting package \citep{Houck00} and following the method given by \citet{Starling07}, having the advantage that no model for the X-ray data needed to be assumed a priori. Furthermore, as all the data are flux-calibrated, no additional offset were needed. We fit the SED using models consisting of an absorbed power law or absorbed broken power law with slopes free or fixed to a slope difference of $\Delta \beta = 0.5$, as expected for a cooling break \citep{Sari98}. The latter model is referred to as ``tied broken power law'', since the spectral slope at lower energies than the break energy $E_{\rm bk}$, $\beta_1$, is tied to the spectral slope at higher energies, $\beta_2$ ($\beta_1=\beta_2-0.5$). The intrinsic optical extinction was modelled with either Milky Way (MW), Large Magellanic Cloud (LMC) or Small Magellanic Cloud (SMC) extinction curves, as parametrized by \citet{Pei92}. The X-ray absorption is modelled with the \begin{small}XSPEC\end{small} model \textit{zphabs}, assuming Solar metallicity. We also tested the model for a lower metallicity. The results for each independent epoch are presented in the appendix, Table \ref{isis_table}. We note that the flaring activity, also reported by \citet{Updike08}, significantly affects the SED modelling at early epochs. 

\begin{figure*}
\includegraphics[width=9cm,angle=-90]{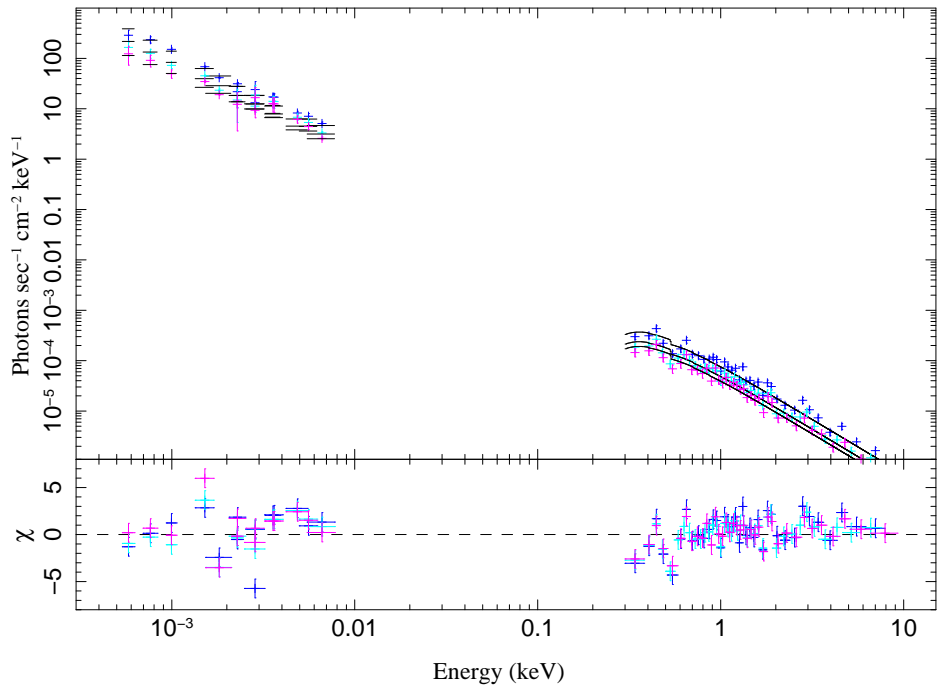}
\caption{Broken power-law fit to the epoch III (blue), IV (cyan) and V (magenta) NIR-to-X-ray SED, with LMC dust extinction and tied parameters between the epochs.}
\label{fig:sed_pl tied}
\end{figure*}

We also explored the SED evolution by tying some parameters between the epochs, where they are not expected to change in time. The tied parameters are the host $E(B-V)$ and $N$(H), since we do not expect the host extinction and absorption to vary at two days after the burst, and the spectral slope at higher energies, $\beta_2$, since we detected no significant X-ray spectral evolution (Sec. \ref{sec x-rays}). The spectral slope at lower energies, $\beta_1$, and $E_{\rm bk}$ were left free to vary between the epochs. We excluded epoch I and epoch II from the SED fit tied between the epochs to avoid the flaring activity at early times, see Fig. \ref{fig:alllc}. These joint fits are listed in Table \ref{isis_table tied} and Fig. \ref{fig:sed_pl tied} shows the best fitting joint model to the last 3 epochs. We discuss the best fitting model and a comparison with the results of broadband SED modelling in Sec. \ref{sec discussion}.

\begin{table*}
\centering
\caption{NIR-to-X-ray SED modelling with a single power law (PL), a broken power law (BPL) and a tied broken power law (TIED BPL), where the spectral slopes are tied to differ by $\Delta\beta\equiv0.5$, as expected for a synchrotron cooling break \citep{Sari98}. Each model is jointly fit to the SED at epoch III, IV and V (1.9, 2.5 and 2.9 days after the trigger) by tying some parameters between the epochs, i.e. $E(B-V)$, $N$(H) and $\beta_2$ are allowed to vary but not between the epochs. The best fitting model (underlined) is a broken power law with $\beta_2=1.18\pm0.01$ and a break energy $E_{\rm bk}=7.6^{+4.0}_{-1.8},\,5.3^{+1.2}_{-1.0},\,4.3^{+0.9}_{-0.6}$ eV, in the UV. The $N$(H) is consistent with the Galactic value, while a low dust content $E(B-V)=0.016\pm0.007$ mag can be derived from the SED, equally well described by SMC and LMC extinction curves. The errors refer to a 90\% confidence limit, and are solely statistical error, i.e. they do not include systematic errors.}
\begin{tabular}{l | l l l l l l l r }
\hline \hline
\rule[-0.2cm]{0mm}{0.6cm}
  Ext. &  Model &  $\chi^2_\nu$ [dof] & $E(B-V)$ & $N$(H) & $\beta_1$ & $\beta_2$ & $E_{\rm bk}$ & Epoch \\
\rule[-0.2cm]{0mm}{0.6cm}
 type &        &                &  (mag) & ($10^{22}$ cm$^{-2}$)& &                &   (eV)      \\ 
\hline
\rule[-0.0cm]{0mm}{0.4cm}

           &    PL      &   4.06 [148]    &   $<0.0018$   &   $<1.3\times 10^{-6}$   & $1.083\pm0.008$   &   ---------    &    ---------   & III \\
          &             &         &                         &            &   $1.091\pm0.009$                        &   ---------        &     ---------            & IV \\
\rule[-0.2cm]{0mm}{0.4cm}
          &             &         &                         &            &   $1.069^{+0.010}_{-0.009}$  &  ---------         &     ---------            & V \\
   
          &    \underline{BPL}     &    2.85 [146]   & $0.015\pm0.007$ & $<1.2\times 10^{-6}$  &  $0.828\pm0.065$   &  $1.18\pm0.01$    &  $7.5^{+3.7}_{-1.8}$          & III \\
 \underline{SMC}  &             &         &                         &            &   $0.72^{+0.08}_{-0.09}$  &           &    $5.4^{+1.3}_{-0.9}$     & IV \\
          \rule[-0.2cm]{0mm}{0.4cm}
          &             &         &                         &            &   $0.48^{+0.12}_{-0.13}$  &           &    $4.5^{+0.9}_{-0.7}$      & V \\
           
          & TIED BPL &    2.96 [147]   & $0.017\pm0.006$  & $<1.2\times 10^{-6}$  &   $\beta_2-0.5$   & $1.18\pm0.01$    &    $5.1^{+0.9}_{-0.7}$         & III \\
          &             &         &                         &            &      &           &   $5.1^{+0.8}_{-0.7}$   & IV \\
          \rule[-0.2cm]{0mm}{0.4cm}
          &             &         &                         &            &      &           &   $6.0^{+1.1}_{-0.8}$   & V \\
  
\hline
\rule[-0.0cm]{0mm}{0.4cm}

          &    PL      &    4.06 [148]    &  $<0.0018$     &  $<1.3\times 10^{-6}$         & $1.083\pm0.008$   &   ---------        &     ---------            & III \\
          &             &         &                         &            &  $1.091\pm0.009$   &   ---------        &     ---------            & IV \\
\rule[-0.2cm]{0mm}{0.4cm}
          &             &         &                         &            &   $1.069^{+0.010}_{-0.009}$  &  ---------         &     ---------            & V \\

          &    \underline{BPL}     &    2.84 [146]   & $0.016\pm0.007$ &  $<1.3\times 10^{-6}$ & $0.835^{+0.065}_{-0.063}$    & $1.18\pm0.01$     &            $7.6^{+4.0}_{-1.8}$ & III \\
 \underline{LMC}  &             &         &                         &            &   $0.72^{+0.08}_{-0.09}$  &           &    $5.3^{+1.2}_{-1.0}$     & IV \\
          \rule[-0.2cm]{0mm}{0.4cm}
          &             &         &                         &            &   $0.45^{+0.12}_{-0.13}$  &           &   $4.3^{+0.9}_{-0.6}$      & V \\
   
          & TIED BPL &     2.97 [147]   & $0.016\pm0.006$  & $<1.2\times 10^{-6}$  &   $\beta_2-0.5$   &  $1.18\pm0.01$   &     $5.2^{+0.9}_{-0.7}$        & III \\
          &             &         &                         &            &      &           &  $5.1^{+0.8}_{-0.7}$    & IV \\
          \rule[-0.2cm]{0mm}{0.4cm}
          &             &         &                         &            &      &           &   $6.0^{+1.2}_{-0.8}$   & V \\
  
\hline
\rule[-0.0cm]{0mm}{0.4cm}

          &    PL      &    4.06 [148]    &  $<0.0017$     &   $<1.3\times 10^{-6}$     &  $1.083\pm0.008$  &   ---------        &     ---------            & III \\
          &             &         &                         &            &  $1.091\pm0.009$   &   ---------        &     ---------            & IV \\
\rule[-0.2cm]{0mm}{0.4cm}
          &             &         &                         &            &  $1.069^{+0.010}_{-0.009}$   &  ---------         &     ---------            & V \\
      
          &    BPL     &     2.99 [146]   &$0.008\pm0.007$  &  $<1.2\times 10^{-6}$ &  $1.089^{+0.010}_{-0.009}$   &   $1.184\pm0.016$   &            & III \\
 MW  &             &         &                         &            &   $0.71^{+0.07}_{-0.08}$  &           &    $6.1^{+1.7}_{-1.0}$     & IV \\
          \rule[-0.2cm]{0mm}{0.4cm}
          &           &         &                         &            &   $0.50^{+0.10}_{-0.12}$   &          &     $<5$    & V \\

          & TIED BPL &      4.14 [147]   & $<0.0016$  &  $<1.2\times 10^{-6}$ &   $\beta_2-0.5$   &  $1.581^{+0.006}_{-0.005}$   &    $       10.00^{+0.00}_{-6.56} \times 10^3$         & III \\
          &             &         &                         &            &      &           &  $3.17^{+6.83}_{-0.77}\times 10^3$   & IV \\
          \rule[-0.2cm]{0mm}{0.4cm}
          &             &         &                         &            &      &           &  $7.58^{+2.42}_{-4.30}\times 10^3$   & V \\

\hline \hline
\end{tabular}
\label{isis_table tied}
\end{table*}

\label{sec mod blast}

\section{Discussion}
\label{sec discussion}

\subsection{The spectral energy distribution}
\label{sec sed discussion}

The jointly fit epoch III--V NIR-to-X-ray SED of \grb{} is best described by a broken power law with $\beta_2=1.18\pm0.01$ (tied between the epochs), and $\beta_1=0.835^{+0.065}_{-0.063},\,0.72^{+0.08}_{-0.09},\,0.45^{+0.12}_{-0.13}$ for epoch III, IV and V respectively ($\chi^2_\nu=2.84$ for 146 degrees of freedom, dof, see Table \ref{isis_table tied}). The break energy, $E_{\rm bk}$ is located in the UV and there is a marginal hint for its evolution\footnote{Consistent with a time evolution of the synchrotron cooling frequency as $\nu_c \propto t^{-1/2}$ expected for a blast wave expanding in a constant ISM medium in the slow cooling regime \citep[$\nu_m<\nu_c$, e.g.,][]{Sari98}. However, the evolution is not significant, since fitting a constant to the three $E_{\rm bk}$ values provides $\chi^2=1.4$ (for 2 dof).} in time towards longer wavelengths: $\lambda_{\rm bk} = 1632^{+506}_{-562},\,2340^{+544}_{-432},\,2885^{+468}_{-499}$ \AA{} from epoch III, IV and V. The quoted errors refer to 90\% confidence. The spectral slope differences suggest that the spectral break is a synchrotron cooling break, and a fit with a tied broken power law ($\Delta\beta=0.5$) provides an adequate fit indeed ($\chi^2_\nu=2.97$ for 147 dof) and the same $\beta_2=1.18\pm0.01$. The free broken power law is slightly preferred with respect to the tied broken power law, but the latter cannot formally be excluded (F-test null probability of 0.62). Regardless of the extinction model used, a broken power law always gives a significantly better fit than a single power law from the NIR to X-ray regime. The F-test probability of the broken power-law over the power-law fit improvement being obtained by chance is between $10^{-12}$ and $7\times 10^{-11}$ with all extinction laws in the epochs III-V joint fits, confirming the significance of the improvement.

 The dust extinction $A_V=0.05\pm0.02$ mag \citep[for $R_V=A_V/E(B-V)=3.16$ in the LMC,][]{Pei92} is best modelled with a LMC or SMC extinction curve, suggesting a low metallicity GRB absorber\footnote{[Fe/H] $=-1.25\pm0.01$ ($\sim -1.15$) in SMC (LMC) star clusters \citep{Cioni09}, while the Milky Way metallicity distribution peaks around Solar values both in the bulge \citep[e.g.,][]{Zoccali03} and in the thin disk \citep{Tiede99}, covering the majority of the stellar population.}. We note that the free broken power law with MW-type extinction curve cannot be formally excluded by the fit ($\chi^2_\nu=2.99$ for 146 dof). However, the very low dust content implied by this solution ($E(B-V)=0.008\pm0.007$) also suggests a low metallicity for the absorbing gas \citep[e.g.,][]{Vladilo98}.

 $N$(H) cannot be constrained, i.e., it is consistent with the Galactic value and with the large error of the X-ray-only $N$(H) $=15.5^{+13.9}_{-14.9}\times 10^{20}$ cm$^{-2}$. This $N$(H) falls at the low end of the $N$(H) distribution of GRB afterglows \citep[e.g.,][]{Evans09,Fynbo09,Campana10}, and $A_V$ also falls at the low end of the dust extinction distribution \citep[e.g.,][]{Kann10,Schady11,Zafar11}. The dust to gas ratio $N$(H)$/A_V=3.1^{+2.7}_{-3.0}\times 10^{22}$ cm$^{-2}$ mag$^{-1}$ covers most of the GRB $N$(H)$/A_V$ distribution at its low end \citep[see][]{Zafar11} and is similar to GRB\, 980329 \citep{Starling07}. In fact, one in every six {\it Swift} XRT X-ray afterglows have unconstrained excess $N$(H) \citep{Evans09}, given the difficulty of measuring low host galaxy $N$(H) against higher Galactic $N$(H) and its uncertainties.

At earlier epochs I and II the afterglow SED is affected by flaring activity. This is most evident in the $V$ and $R$ bands where epochs I and II coincide with the rise and decay of a flare or bump in the light curves (Fig. \ref{fig:alllc}). For completeness we report individual fits to these early epochs in the appendix. Our NIR-to-X-ray SED results are in overall agreement with the single epoch SED at 4.3 days previously analysed by \citet{Updike08}. In particular, we confirm the presence of a spectral break. This break was only tentatively found by \citet{Updike08}, because of its sensitivity to the NIR continuum level, based on a large extrapolation of solely two data-points in the $K$ band. In our dataset, 6 additional NIR data-points were added, i.e. $K_s, J, H$ from the SMARTS telescope \citep{Kann10} and $K_s, J, H$ from the PAIRITEL telescope \citep[from GCN,][]{Chandra08}. Moreover, no extrapolations were needed for the NIR light-curve from epoch III to epoch V of our SED fit. Finally, the significant improvement from a single power law SED fit to a broken power law strongly favours the presence of the spectral break.

\subsection{The broadband picture}
\label{sec blast discussion}

To determine the full set of physical parameters of a GRB afterglow, it is vital to have good temporal and spectral coverage from X-ray to radio frequencies. \grb{} is a good candidate for this kind of study, in particular because of its coverage at centimetre wavelengths. With well sampled light curves at multiple radio observing frequencies one can determine the evolution of the peak flux, the associated peak frequency $\nu_{\rm{m}}$, and the synchrotron self-absorption frequency $\nu_{\rm{a}}$. Together with the cooling frequency $\nu_{\rm{c}}$, the electron energy distribution index $p$, and the jet-break time, all the micro- and macrophysical parameters of the GRB jet and its surroundings can be pinned down. For \grb{} this has been done by \citet{Chandra08} for part of the total available data set. \citet{Updike08} have focused their modelling efforts mainly on the optical and X-ray regimes, as we have done in Sec. \ref{sec sed discussion}. There are some inconsistencies between the different modelling results, and complications in interpreting the full data set, which we will discuss here.

The main issue is the position of the cooling break in the broadband spectrum. Our SED fits show that there is a spectral break between the optical and X-ray bands. The spectral slope differences between the two bands indicate that this break can be interpreted as the cooling break. As pointed out by \citet{Updike08}, it is difficult to form a coherent picture of these spectral slopes and the temporal slopes observed at optical and X-ray frequencies within the standard afterglow framework, because not all the so-called closure relations \citep[e.g.,][]{Zhang04} can be satisfied. \citet{Updike08} suggested a solution for this issue by adopting that the NIR-to-X-ray SED is in fact a single power law, and they concluded that all those observing bands are in between $\nu_{\rm{m}}$ and $\nu_{\rm{c}}$ and that the circumburst medium is homogeneous. This solution is not satisfactory given that we confirm the presence of a cooling break in the NIR-to-X-ray SED, similar to what was found and tentatively discarded by \citet{Updike08}. 

A complicating factor is the flaring behaviour of \grb{}, which makes reliable determinations of temporal slopes difficult. Our SEDs have been constructed at times when the flaring has ceased, but leaving out the flaring parts of the optical and X-ray light curves gives a relatively short lever arm for determining the temporal slopes, in turn giving rise to large statistical and systematic uncertainties. This could account for the fact that the standard closure relations for $\nu_{\rm{c}}$ in between the optical and X-ray bands do not seem to work. For example, it is mentioned in \citet{Updike08} that the optical temporal slopes are consistent with the spectral slopes if the circumburst medium is structured like a stellar wind, but the X-ray temporal slope is inconsistent. The latter could be due to uncertainties caused by the flaring of the source. The bottom line is that in the case of \grb{} the NIR-to-X-ray SED fits seem to give the most reliable measurement of the value of $\nu_{\rm{c}}$. 

The broadband SED fitting performed by \citet{Chandra08}, including the centimetre wavelength radio data, leads to different conclusions. They find that the optical and X-ray regimes lie on the same power-law segment of the synchrotron spectrum, but above both $\nu_{\rm{m}}$ and $\nu_{\rm{c}}$. In fact, one of the outcomes of their modelling is that this afterglow is in the fast cooling regime, i.e. $\nu_{\rm{c}}<\nu_{\rm{m}}$, for about a week, which has not been observed in any GRB afterglow before. The values for $p$ they find result in optical and X-ray spectral indices of $\sim1.1-1.2$. These indices are consistent with the X-ray spectral index we find in our SED fitting, but inconsistent with the shallower optical indices we find, and also the indices obtained by \citet{Updike08} for NIR-optical SED fits. The uncharacteristically low value for $\nu_{\rm{c}}$ is necessary to describe the optical and X-ray light curves together with the bright and long-lasting radio afterglow. If $\nu_{\rm{c}}$ would be at higher frequencies, there is no possible combination of $\nu_{\rm{a}}$, $\nu_{\rm{m}}$ and the peak flux to give a satisfactory fit of the radio data. Although the modeling performed by \citet{Chandra08} is thorough, the basic assumptions are fairly simplified. To get to a coherent solution for all the broadband data, a more complicated picture may have to be invoked, like a double jet as has been suggested for the broadband afterglow of e.g. GRB\,030329 \citep{Berger03}. Another possibility is that we need a more detailed description of the afterglow physics, in particular the effects of synchrotron self-absorption on the jet-break time at radio wavelengths and the associated changes this makes to radio light curves \citep{VanEerten11}. These different, more complicated scenarios for satisfactorily describing the full data set of \grb{} are beyond the scope of this work, but will be explored and presented in a future paper.

While the broadband modelling needs further study, we can still rely on the NIR-to-X-ray SED, as it is based on fewer assumptions. In particular, we found a low $N$(H) and a low SMC/LMC-type dust content, suggesting a low metallicity environment.

\subsection{The absorber environment}
\label{sec absorber discussion}

\begin{figure*}
\includegraphics[width=130mm,angle=0]{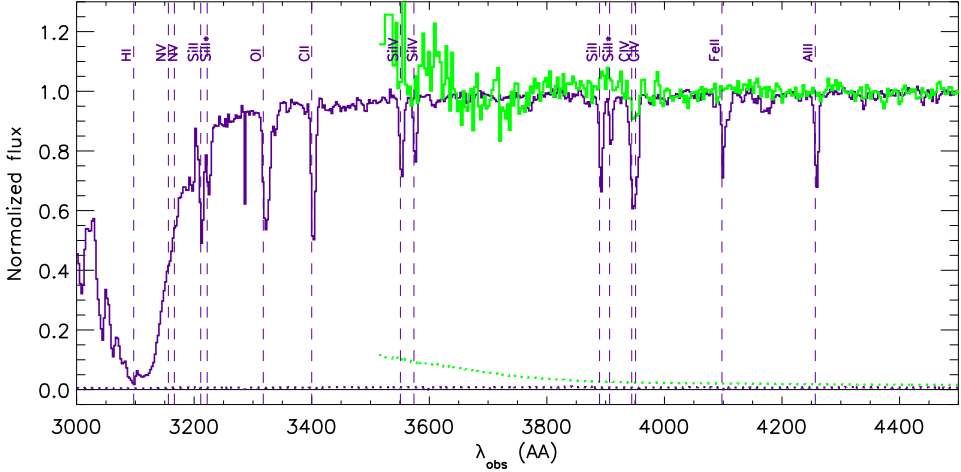}
\caption{The high S/N GRB composite spectrum (dark) from \citet{Christensen11} shifted to $z=1.5477$ and the observed \grb{} FORS spectrum (light) cut at S/N $>10$. The error spectra are shown by the dotted lines. A firm upper limit of $z \lesssim 1.9$ can be placed on the \grb{} redshift from the lack of \lya{} forest lines in the afterglow spectrum.}
\label{fig composite}
\end{figure*}

While the broadband SED depends on the afterglow physics and the circumburst environment, the absorption in the optical spectrum probes the gas along the line of sight. By studying the nature of the absorber we can derive information on the GRB site, as we will discuss here.

A crucial issue that needs to be addressed is whether or not the \mgii{} absorbing system observed in the afterglow spectrum is associated with the host galaxy of \grb{}. While a host galaxy could, in principle, be in the background at a larger redshift, we note that \lya{} absorption at $z \gtrsim 1.9$\footnote{Corresponding to the bluest observable wavelength $\lambda=3515$ \AA{} in the spectrum cut where S/N$\sim10$} would be observable in our high S/N FORS spectra, but no signature of such system was detected. Figure \ref{fig composite} shows how the observed \grb{} FORS spectrum compares to the GRB composite spectrum from \citet{Christensen11} at the same redshift $z=1.5477$. A $1.5 \lesssim z \lesssim 1.9$ host galaxy could be possible but no lines are detected at such redshift. Thus, we consider $z=1.5477$ to be the host galaxy redshift. A tentative \lya{} absorption at this redshift was reported by \citet{Updike08}, for which $\log N$(\hi{}) $<20.3$ was argued. We note that high-ionization \nv{} $\lambda\lambda$ 1238, 1242 \AA{} lines are not in our wavelength reach. No spectroscopic observation has been attempted to investigate the two candidate blue hosts that \citet{Cenko08} indicated at a large offset from the GRB position so far. Below we will discuss the possible scenarios.

\begin{figure}
\includegraphics[width=8.5cm]{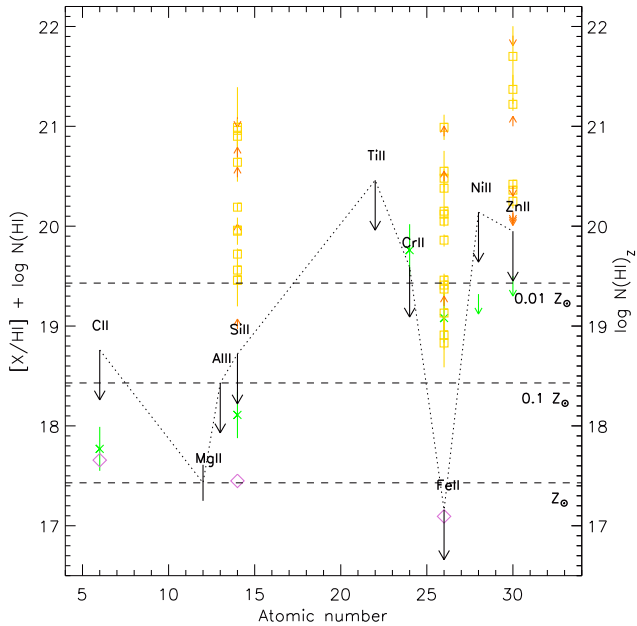}
\caption{Abundances and 3$\sigma$ detection limits of a selection of low ionization species in the \grb{} afterglow spectrum (black bar and arrows) from Table \ref{tab_lines} and Table \ref{tab_ul}. The ordinate combines the metallicity and the \hi{} column density, basically a measure of the metal abundances with respect to Solar, in order to show the possible ranges of metallicity and $N$(\hi{}) in the absorber. The horizontal dashed lines correspond to the neutral hydrogen column density for a given metallicity, log N(\hi{})$_Z$, with respect to the \mgii{} detection, assuming no depletion of Mg into dust. Light-coloured squares and show the metal abundances observed in GRB absorbers \citep{Schady11} and the crosses refer to the GRB\,080310 measurements (De Cia et~al., in prep.). The column density expectations (diamonds) for the measured \mgii{} are derived from the correlations observed in weak Mg absorbers \citep{Narayanan08}.}
\label{fig metul}
\end{figure}

The metallicity along the \grb{} line of sight cannot simply be constrained from the measured column densities or upper limits, since no \lya{} is available in our FORS spectra, given the redshift of this afterglow. One way to bypass the lack of $N$(\hi{}) measurement is plotting [X/\hi{}] $+$ log\,$N$(\hi{}) $=$ log\,$N$(X) $-$ log\,$N$(X)$_\odot +$ log\,$N$(\hi{})$_\odot$, as in Fig. \ref{fig metul}. Despite the degeneracy between \hi{} column density and metallicity, we can use the \mgii{} detection and the upper limits to explore the parameter space and at least limit the possible $N$(\hi{}) and [X/\hi{}]. In particular, assuming Solar metallicity and no Mg depletion into dust, we derive log\,$N$(\hi{}) $=17.43$. If the metallicity is 1/100 Solar then log\,$N$(\hi{}) $=19.43$. A much lower metallicity is very hard to justify, while a super-Solar metallicity with lower $N$(\hi{}) is possible. Hence, we can reduce the possibilities for the \grb{} absorber to be either \textit{i)} a sub-Lyman-Limit System \citep[sub-LLS, log\,$N$(\hi{}) $<17.3$,][]{Rauch98} with $Z>1.3\,Z_\odot$, or \textit{ii)} an LLS \citep[$17.3<$ log\,$N$(\hi{}) $<19.0$,][]{Peroux03} with $0.03\,Z_\odot<Z<1.3\,Z_\odot$, or \textit{iii)} a sub-DLA ($19.0<$ log\,$N$(\hi{}) $<20.3$) with $0.001\,Z_\odot<Z<0.03\,Z_\odot$. Below we discuss these three different scenarios.

\textit{i) A super-Solar $Z$ sub-LLS.} This scenario is supported by the low \mgii{} $EW$ of \grb{}, since the majority of weak \mgii{} ($EW<0.3$ \AA{}) absorbers in quasar line of sights are associated with sub-LLS with super-Solar metallicity \citep{Churchill99}. This association is based on the number density consistency between the two populations and a \begin{small}CLOUDY\end{small} photo-ionization model that predicts log N(\hi{})~$<17.3$ for weak \mgii{}. Given the observed \mgii{} column density in \grb{}, the \cii{}, \siii{} and \feii{} expectation for weak \mgii{} systems derived by \citet{Narayanan08} (diamonds in Fig. \ref{fig metul}), are consistent with our upper limits. These systems could trace either interstellar gas expelled from star forming regions into the halo, or small clouds in dwarf galaxies or intergalactic star-forming structures \citep{Churchill99}. The least likely possibility is the GRB explosion site in the halo. \citet{Cenko08} already pointed out the difficulty of explaining how to ``kick" the massive GRB progenitor star all the way out to the galaxy halo ($\sim27$ kpc). A halo in-situ formation of the GRB would require halo gas at $\sim27$ kpc and $\sim46$ kpc from two interacting galaxies to trigger massive star formation, the GRB progenitor producing high density stellar winds to ``fuel'' the extremely bright afterglow \citep[$n\approx50$ cm$^{-3}$,][]{Chandra08}. One other possibility is that the GRB exploded in a dwarf galaxy. \citet{Ellison04} studied the \mgii{} systems coherence length, indicating that the weak \mgii{} absorbers have sizes of 1.5--4.4 kpc and they could represent a distinct population of smaller galaxies. However, the low content of SMC/LMC-type dust derived from the \grb{} afterglow SED suggests a low metallicity of the host galaxy, disfavouring the super-Solar $Z$ sub-LLS scenario.

\textit{ii) A $0.03\,Z_\odot<Z<1.3\,Z_\odot$ LLS.} Although GRB absorbers show, on average, high \hi{} column densities \citep[e.g.,][]{Jakobsson06}, an LLS is not an exotic environment for a GRB, since a low N(\hi) may be a line of sight effect. In principle, the burst radiation can travel along the line of sight, within the host galaxy or from the galaxy outskirts, without being absorbed by dense gas clouds. This scenario is supported by the observation of a few other GRB LLSs, namely GRB\,050908 \citep[log\,$N$(\hi{}) $=17.60\pm0.10$:][]{Fugazza05,Fynbo09}, GRB\,060124 \citep[log\,$N$(\hi{}) $=18.5\pm0.5$:][]{Prochaska06b,Fynbo09}, GRB\,060607A \citep[log\,$N$(\hi{}) $\sim16.8$:][]{Ledoux06,Prochaska08b}, GRB\,080310 (log\,$N$(\hi{}) $=18.70\pm0.10$: De Cia et~al., in preparation) and GRB\,090426 \citep[log\,$N$(\hi{}) $=18.70^{+0.1}_{-0.2}$:][]{Thone11}, where \lya{} variability was induced by \hi{} ionization. The low \hi{} content of all these systems cannot be explained by ionization alone, as the burst radiation can totally ionize \hi{} only out to 200--300 pc\footnote{For much brighter afterglows, these distances can significantly grow, see our discussion below.} from the burst in LLSs \citep[using the bright GRB\,050730 afterglow light curve,][]{Ledoux09}. A low \hi{} could arise if the GRB is not too deeply embedded in its host galaxy. Such a possibility, which is certainly expected in some cases, will be further investigated later on in this Section.

The equivalent hydrogen column density measured from the soft X-ray absorption can also provide important information on the gas along the line of sight, with the advantage of not depending on the \hi{} ionization. Indeed, $N$(\hi{}) and $N$(H) do not correlate \citep{Watson07,Campana10}. The difference between a low neutral and a high total hydrogen column densities can be explained if most of the X-ray absorbing gas lies closer to the burst, where the gas is highly ionized \citep{Schady11}. In this scenario, the \grb{} could still have been surrounded by the high density gas required to power such a bright afterglow.

\textit{iii) A $0.001\,Z_\odot<Z<0.03\,Z_\odot$ sub-DLA.} This metallicity range is low, but not unusual for GRB absorbers. For instance, most of the GRB absorbers in the VLT/UVES sample have $0.006\,Z_\odot<Z<0.05\,Z_\odot$, despite the small number statistics \citep{Ledoux09}, although their \hi{} column densities are generally higher.

From the above analysis, scenarios \textit{ii)} and partly \textit{iii)} are favoured, i.e., the \grb{} environment is consistent with other GRBs. Below we test this further, by comparing the \grb{} properties with the GRB population.

\begin{figure}
\includegraphics[width=88mm,angle=0]{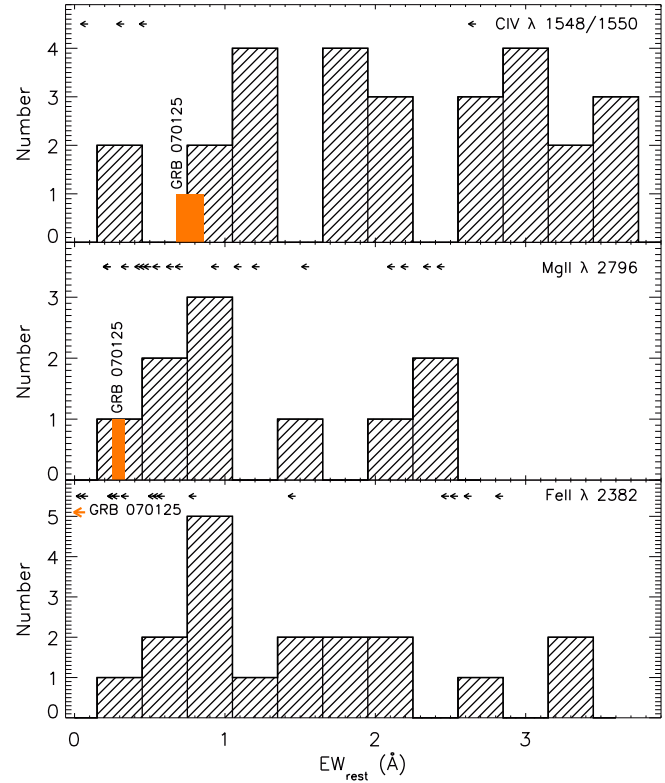}
\caption{The \civ{} $\lambda$ 1548/1550, \mgii{} $\lambda$ 2796 and \feii{} $\lambda$ 2382 restframe $EW$ distributions for GRB absorbers. The $EW$ and $3\sigma$ detection limits are taken from de Ugarte Postigo et~al. (in prep.) and based on \citet{Fynbo09}. The highlighted \grb{} $EW$s fall at the low end of the $EW$ distribution.}
\label{fig ewhist}
\end{figure}

In Fig. \ref{fig metul} we compare the measured metal column densities and upper limits for \grb{} with those observed in other GRB afterglows (squares and arrows), as collected by \citet{Schady11}. The metal column densities in \grb{} are generally lower than measured in other GRBs. This suggests that the \grb{} absorber has either a lower metallicity or a lower \hi{} column density with respect to most GRBs. However, this comparison sample is limited, since the column densities often cannot be constrained, due to the low resolution of the spectra, and more complete $EW$ distributions should be used. The $EW$ of \civ{} $\lambda$ 1549 in \grb{} is $EW_{\textrm{rest}}=0.77\pm0.09$ \AA{}, much lower than in the GRB composite spectrum \citep[$EW_{\textrm{rest}}=2.18\pm0.03$ \AA{}][]{Christensen11} and in the low \hi{} afterglow sub-sample \citep[$EW_{\textrm{rest}}\sim2.8$ \AA{},][]{Thone11}. Nevertheless, the observed \civ{} $\lambda$ 1549 $EW$ and the \siii{} $\lambda$ 1526 upper limit are consistent with the $EW$ correlation found in \citet{Fynbo09}. This suggests that the \grb{} absorber fits in the GRB population, rather than being an outlier. In Fig. \ref{fig ewhist} we compare the \civ{} $\lambda\lambda$ 1548, 1550, \mgii{} $\lambda$ 2796 $EW$s and the deep \feii{} $\lambda$ 2382 $3\sigma$ detection limit for \grb{} with the $EW$ distributions discussed by de Ugarte Postigo et~al. (in preparation) for the whole low resolution spectroscopy afterglow sample of \citet{Fynbo09}. The \grb{} absorber falls in the low end of the $EW$ distribution. Again, this indicates that \grb{} is not a GRB population outlier and that other bursts share similar properties.

%
%
%
 
 \begin{table*}
\caption{Spectral-line poor sample \citep[first group, based on the][sample]{Fynbo09} and related cases \citep[second group, not included in the][sample]{Fynbo09}. $^a$ Afterglow brightness at 1 day after the burst, shifted to $z=1$ \citep{Kann10}. $^b$ $R$, $i$ or $z$ afterglow brightness from the acquisition images of the spectra \citep{Fynbo09}. $^c$ No observations available. $^d$ Possible contribution from the SN or the afterglow itself. [1]  \citet{Savaglio09}; [2] \citet{Chen09}; [3] \citet{Laskar11}; [4] \citet{Dellavalle06}; [5] \citet{Gal-Yam06}; [6] \citet{Covino10}; [7] \citet{Fynbo09}; [8] Cenko et al. (in prep.); [9] \citet{Perley08b}; [10] \citet{Sparre11}.} 
\begin{tabular}{ l | c c c c c}
\hline \hline
\rule[-0.2cm]{0mm}{0.6cm}
      GRB        &   $z$    &  $R_c^a$ ($\rm Mag_{acq}$)$^b$ & Underlying host & Notes & References\\ 
\rule[-0.2cm]{0mm}{0.6cm}
                    &             &   mag            & absolute (apparent) magnitude  &  \\
\hline
\rule[-0.0cm]{0mm}{0.4cm}
060607A     & 3.07  &          (14.0)           & $M_{AB}(780nm)>-20.14$ & $\log N$(\hi{}) $=16.85\pm0.10$ & [1], [2], [3]\\                  
060614       & 0.13   &          (19.8)         & $M_B =-15.5$   & Host $\log M/M_\odot=7.95$ & [1], [4], [5]\\
060708       & 1.92   &           (22.9)         & ? $^c$ & Photometric redshift (low S/N) & \\
060908       &1.88    & $22.5\pm0.5$     & ($R\sim25.6$) & & [1], [6]\\ 
061021       & 0.35   &            (20.5)         & Detected &  Redshift from host and afterglow & [7]\\ 
070419A     & 0.97  &  $23.1\pm0.5$    &   ?  &  &\\ 
       \rule[-0.2cm]{0mm}{0.4cm}
070125       & 1.55  &  $17.5\pm0.1$     &     $M(780nm)> −18.5$ & & [8]\\ 
\hline
\rule[-0.0cm]{0mm}{0.4cm}
071003      & 1.60   &  $17.7\pm0.1$   &  $M(K')>-22.2$ & Halo environment claimed & [9]\\ 
       \rule[-0.2cm]{0mm}{0.4cm}
101219B    & 0.55   &   (19.8)   &  ($r = 23.7\pm0.2^d$) & Associated SN & [10]\\ 
 
\hline \hline
\end{tabular}

\label{tab_slp}
\end{table*}

Furthermore, we selected two afterglow sub-samples that may potentially be similar to \grb{}: \textit{\textit{a)}} the spectral-line poor, i.e., afterglows with no more than two species in absorption at the GRB redshift, selected from \citet{Fynbo09} and \textit{\textit{b)}} afterglows with low equivalent hydrogen column density $N$(H) measured from the X-rays, $N$(H) $<5\times 10^{21}$ cm$^{-2}$, within $3\sigma$. $N$(H) and $A_V$ were collected from \citet{Zafar11} when possible, otherwise the $N$(H) from \citet{Evans09} and the $A_V$ from \citet{Kann10} were used. The $N(\hi{})$ were taken from \citet{Fynbo09} and the $EW$s from de Ugarte Postigo et~al. (in prep.), with the exception of GRB\,060708 and GRB\,080310, whose metal $EW$s are measured here. We exclude the low S/N spectra from the spectral-line poor sample by applying a cut in the \mgii{} $\lambda$ 2796 detection limit of $\sigma_{EW}<0.5$. The low S/N spectra are also excluded from the $EW$ constraints of the low $N$(H) sub-sample. Six afterglows fall in our spectral-line poor sub-sample, not including \grb{}, listed in Table \ref{tab_slp}, while fifteen are in the low $N$(H) sub-sample. 

We note that the spectral-line poor sub-sample may suffer from selection effects. First, afterglows with low redshift ($z\leq0.5$) may more easily be included in this sub-sample because of the shortage of observable transitions (other than the \mgii{} $\lambda\lambda$ 2796, 2803 doublet) in the spectral region covered by typical optical spectrographs. However, only two GRBs have such low $z$ in Table \ref{tab_slp}. Second, the S/N requirement may bias the sub-sample towards brighter bursts. This is irrelevant when comparing the sub-sample properties with \grb{}, the latter being very bright itself, but should be kept in mind when considering the spectral-line poor sub-sample as a population. A more extended and complete sample with deeper S/N limits is needed to further analyse and possibly exclude these selection effects.

\begin{figure*}
\includegraphics[width=160mm,angle=0]{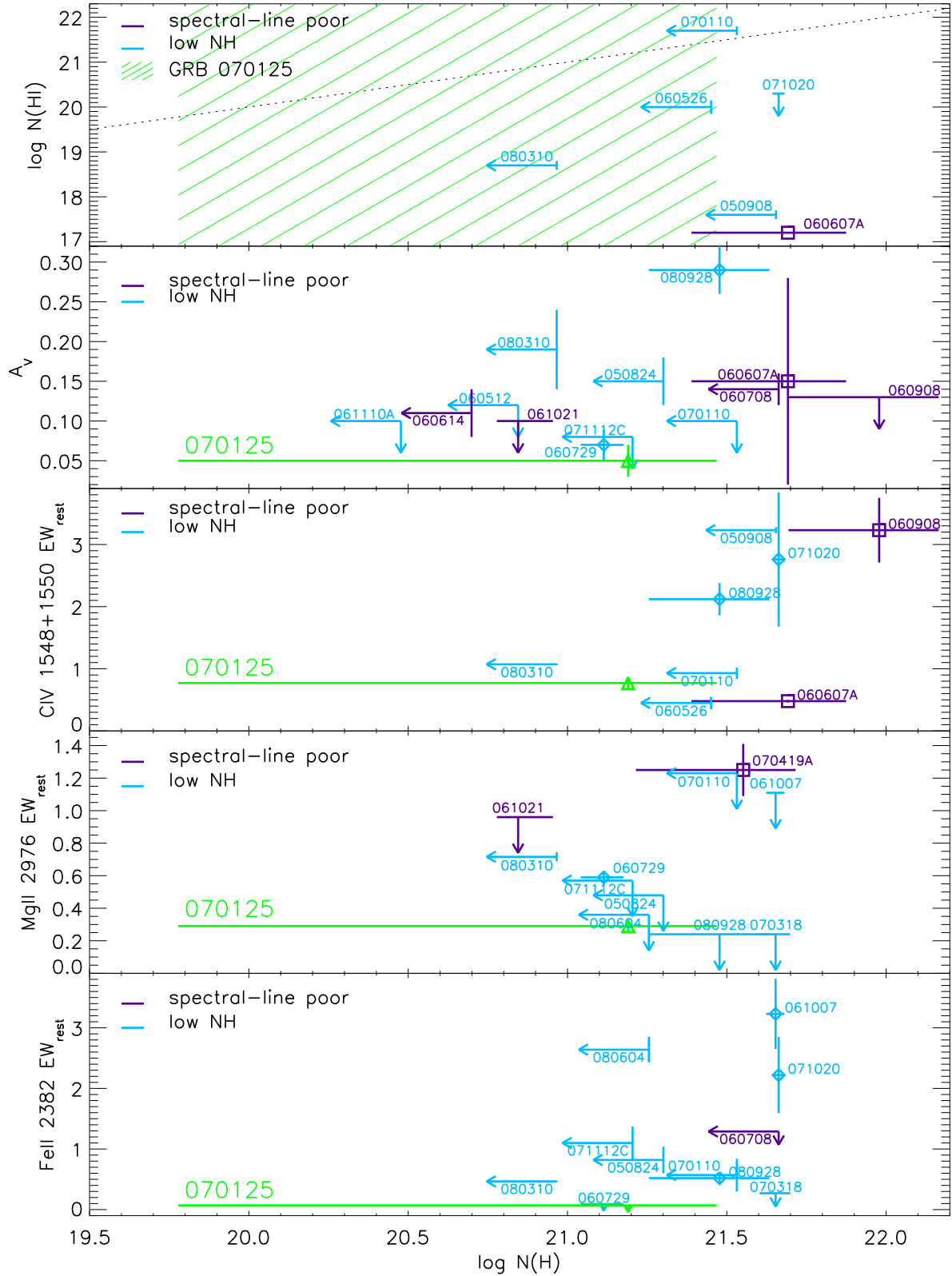}
\caption{The neutral hydrogen column density, dust extinction and restframe $EW$s vs the equivalent hydrogen column density of the spectral-line poor (dark) and the low $N$(H) (light) GRB sub-samples. The \grb{} properties are highlighted. The dotted line in the top panel marks where $N$(H) and $N$(\hi{}) are equal. We do not display the four highest $A_V$ values ($0.34<A_V<0.49$) at the high $N$(H) distribution end, in order to focus on the low $A_V$ and $N$(H) cases most similar to \grb{}. A complete $N$(H) vs $A_V$ distribution is presented in \citet{Zafar11}.}
\label{fig lownh}
\end{figure*}

Fig. \ref{fig lownh} shows a comparison of the \grb{} afterglow absorber properties with the two sub-samples. \grb{} traces the low end of the $N$(H), $A_V$ and $EW$ distributions, confirming the previous finding for the whole sample.  

In general, we notice that most of the spectral-line poor afterglows also show a low $N$(H). The few low $N$(H) afterglows with \lya{} measurements seem to prefer low \hi{} environment, although no conclusion can be drawn yet due to the extremely poor statistics. The same applies to the lack of high $EW$ at constrained low $N$(H). The above suggestions, if confirmed, may indicate a overall low column density environment for both the low-ionization gas (that absorbs the metal lines in the optical spectrum at distances larger than a few hundred pc from the burst) and the highly ionized gas (that absorbs the soft X-rays, in the vicinity of the burst). More observations are needed to investigate whether $N$(H) and $N$(\hi{}) could correlate under special conditions, e.g. due to the effect of a short line of sight within the host galaxy, despite there generally being no such correlation. However, we note that the $N$(H) typically cannot be constrained below $\log N$(H) $<20$--21, due to the uncertainty in the Galactic $N$(H). 

One GRB absorber potentially similar to \grb{} is the recent nearby GRB\,101219B, with redshift $z=0.55$ identified from the \mgii{} $\lambda\lambda$ 2796, 2803 \AA{} and \mgi{} $\lambda$ 2853, the only features in the afterglow spectrum. A spectroscopically confirmed supernova was associated with this GRB, supporting both the low redshift and a massive star progenitor for the GRB. The afterglow spectral shape suggested a very low amount of dust ($A_V<0.1$), while the non-detection of the host galaxy continuum or emission lines indicated that a small galaxy was hosting the GRB, at the faint end of the \citet{Savaglio09} distribution \citep{Sparre11}.

Another case analogous to \grb{} was GRB\,071003, a very bright burst ($R\approx 12$ mag about a minute after trigger), with few and very weak absorption lines in the late time spectrum and no underlying bright host. Given the similarity between the two, GRB\,071003 was also claimed to have occurred in a halo environment \citep{Perley08b}. In both cases, while the high brightness of the afterglow cannot easily be associated with a halo star-forming region, it can naturally contribute to ionize the circumburst medium to larger distances. And perhaps the time delay between the burst and the epoch of the spectra ($\sim1$ day in both cases) could allow the ionization front of the bright afterglow to travel further through the host galaxy. 

\begin{figure}
\includegraphics[width=80mm,angle=0]{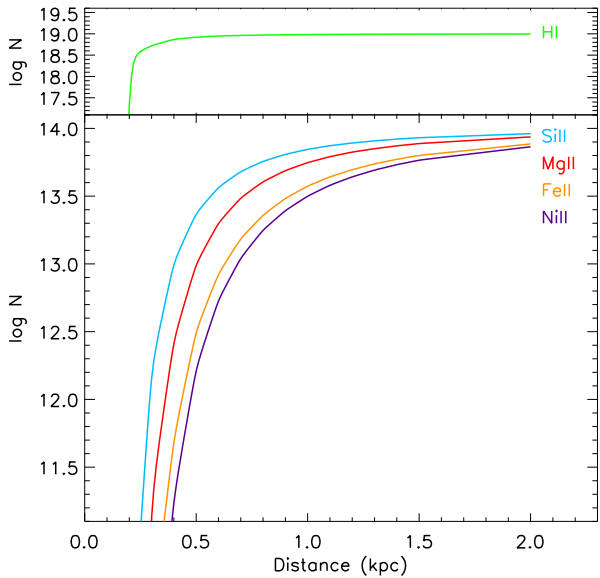}
\caption{The ionization effect of the afterglow radiation on absorbing clouds at a given distance with initial column densities $\log N$(\hi{}) $=19$ and $\log N=14$ for the metals, as observed at 1 day after the burst. See text for details.}
\label{fig dmod}
\end{figure}

In order to investigate this possibility, we performed a simplified simulation of the GRB\,070125 afterglow radiative effect on the surrounding medium with a radiative transfer model, including photo-ionization. A detailed description of the complete model will be discussed in Vreeswijk et~al. (in prep). We use the GRB\,050730 light curve as a template, scaled to the 3.3 times brighter flux of \grb{} \citep[derived from the afterglow brightness at 1 day, if both GRBs would lie at $z=1$, $R_c$(\begin{small}GRB\,050730\end{small}) $-R_c$(\begin{small}GRB\,070125\end{small}) $=1.3 $ mag,][]{Kann10}. We assume a spectral slope $\beta=0.7$, as before the synchrotron cooling, since the most relevant photons are the $UV$ in the first hours after the burst. We let the GRB afterglow radiate through a cloud lying at a distance between 0 and 2 kpc and investigate the variation of the ionic column densities. In general, most of GRB absorbers have been found to lie within this distance range from the burst, in those cases for which  the GRB-to-cloud distance could be estimated from the variability of fine-structure lines in the spectra. Two different sets of initial column densities were chosen: \textit{a)} $N$(\hi{}) $=10^{19}$ cm$^{-2}$ and $N=10^{14}$ cm$^{-2}$ for the metal ions and \textit{b)} $N$(\hi{}) $=10^{20}$ cm$^{-2}$ and $N=10^{15}$ cm$^{-2}$ for the metals. We set the observation times to be at 1 day and 6 hr for comparison (observer frame). The results for case \textit{a)} are displayed in Fig. \ref{fig dmod}. While the higher $N$(\hi{}) shields the metals from being ionized at large distances, in the case of $N$(\hi{}) $=10^{19}$  cm$^{-2}$, most of the metals are strongly ionized out to $\sim0.5$ kpc. In particular, at 0.5 kpc the \mgii{} column density is 10 times lower than the initial value, \feii{} is $\sim$30 times lower than the initial column, \niii{} even lower. The late time of the spectrum (1 day) does not seem to be an important factor, as the comparison with the ionization expected at 6 hr after the burst does not change the result dramatically. In particular \mgii{} and \feii{} are ionized by only a factor of $\sim$2 between 6 and 24 hours, indicating that most of the ionization is happening within the first hours after the burst. At 0.2 kpc all these metals and \hi{} are totally ionized. Thus, we conclude that any cloud within 0.5 kpc would not be visible, in terms of spectral lines of low ionization species. This evidence, together with a GRB location not too deep inside the host galaxy (i.e., within 0.5 kpc) can explain the lack of spectral features in the spectrum of \grb{}. 

The feasibility of this scenario for other bursts can be tested by comparing how many cases similar to \grb{} we expect and how many we actually observe. For a zero order approximation, we calculate the rough probability of a GRB to have \textit{i)} a short line of sight within the host galaxy, \textit{ii)} a low absorbing $N$(\hi{}) and \textit{iii)} a bright afterglow. \textit{i)} The probability of a GRB to be located not deeper than 0.5 kpc within its host galaxy depends on the galaxy size (in one simple dimension it can be simplified as $p(<\rm 0.5\,kpc)\approx0.5\,{\rm kpc}/{\rm size}$. Integrating this probability over a normal distribution of GRB host galaxy sizes centred around the mean size 2.4 kpc \citep[][assuming $\sigma=0.8$ and excluding sizes below 0.5 kpc, to reproduce the observed 0.5--5 kpc size distribution]{Wainwright07} provides a 22\% chance of a GRB line of sight traversing less than 0.5 kpc of its host galaxy. \footnote{This estimate assumes a random distribution of the GRB locations inside the host galaxy. However, the low metallicity requirement for the GRB progenitor formation \citep[e.g.,][]{Hirschi05} would increase the probability of GRBs being located in the outskirts of their hosts, given the typical metallicity gradients.} \textit{ii)} Among the GRB absorbers with detected \lya{} absorption in the \citet{Fynbo09} sample, 15\% are LLS or lower $N$(\hi{}). \textit{iii)} The afterglow brightness distribution, scaled to redshift $z=1$ and at 1 day after the burst, is presented in \citet{Kann10}, providing 16\% of GRBs with $R_c<18$ mag afterglows. A different approach, considering the magnitudes of the acquisition images ($R$, $i$ or $z$-band magnitude) of the afterglow spectra from \citet{Fynbo09}, yields 12\% afterglows with $\rm Mag_{acq}<18$ mag. Combining the three probabilities above we estimate that about 0.5\% of all GRB afterglows are similar to \grb{}, i.e. spectral-line poor and bright. This is somewhat consistent with the observations of \grb{}, GRB\,060607A and including GRB\,071003, the three of them representing 4\% of the afterglows analysed in \citet{Kann10}. We made an approximate estimate of the number of GRBs with a ground-based detection among the {\it Swift} sample (and hence the possibility of being identified as spectral-line poor, low $N$(HI) and optically bright), using the {\it Swift} GRB Table\footnote{\texttt{http://swift.gsfc.nasa.gov/docs/swift/archive/grb{\_}table/}}. This results in 324 GRBs up to 2011 May 1, of which 1\% would mean $\sim$3 sources, again consistent with current observations. If we relax the brightness requirement, about 3\% of all GRB afterglows are expected to have a low $N$(\hi{}) and lie in the outer kiloparsec of their host galaxy along our line of sight. The 9 spectral-line poor afterglows listed in Table \ref{tab_slp} (including GRB\,071003 and GRB\,101219B) represent 12\% of the afterglows of \citet{Kann10}, or 3\% of the \textit{Swift} optical/IR afterglows. It is possible that these spectral-line poor afterglows are the result of a GRB location in the outer regions of their host galaxy and a low $N$(\hi{}). Further investigations are needed to confirm these results.

Finally, we point out that the non-detection of the \grb{} host galaxy in the KeckI/LRIS images down to $R >$ 25.4 mag \citep{Cenko08} is consistent with a $M_{300,\,\rm{AB}}>-18.8$ galaxy at $z=1.5477$ (k-corrected AB magnitude at 300 nm), i.e., a faint (among the 25\% faintest) but still not uncommon GRB host galaxy at that redshift (Schulze et~al. 2011, in preparation). Indeed, GRB hosts are generally faint irregular galaxies \citep{Savaglio09}. Recent \textit{HST} imaging of the field provides deeper upper limits on the host brightness of \textit{F336W} $>$ 26.7 mag (AB) (central wavelength $\lambda_c\sim 334.4$ nm) and \textit{F110W} $>$ 26.4 mag ($\lambda_c\sim 1123.8$ nm), not corrected for foreground extinction (Cenko et~al., in prep.). We then correct for Galactic foreground extinction of 0.26 and 0.05 mag in the F336W and F110W bands, respectively. The former upper limit corresponds to an absolute brightness of $M_{300,\,\rm{AB}} > -18.3\,\rm{mag}$ in the 300-nm rest frame, assuming $F_\nu\propto\nu^{-1/2}$ at that wavelength. The NIR observation in \textit{F110W} can constrain the rest frame optical/NIR brightness of the host galaxy, mostly attributed to late type stars and the old stellar population. The upper limit corresponds to $M>-18.2\,\rm{mag}$ in the 780 nm, assuming $F_\nu\propto\nu^{-1/2}$. As is the case in the UV, this host is among the faintest hosts in the luminosity distributions of \citet{Savaglio09} and \citet{Laskar11}.

\section{Summary and conclusions}
\label{sec conclusions}

We collected all observational data of \grb{} available in the literature, in order to investigate the environmental properties of the burst, and in particular to test the putative halo origin of this burst based on the few weak absorption lines in the spectrum and the absence of an underlying bright host \citep{Cenko08}. 

The NIR-to-X-ray SED at 1.9, 2.5 and 2.9 days after the burst revealed a synchrotron cooling break in the UV. Along the line of sight in the host galaxy, little reddening ($E(B-V) = 0.016\pm0.007$) is caused by a LMC- or SMC-type dust, while a low equivalent hydrogen column density cannot be constrained better than $N$(H)$	=	15.5^{+13.9}_{-14.9} \times 10^{20}$ cm$^{-2}$. The NIR-to-X-ray SED results are inconsistent with broadband SED analyses including radio observations, indicating that the assumptions in modelling efforts for \grb{} so far have probably been too simplified. More complex afterglow models should be explored, in particular to unify the light curves across the broadband spectrum with the NIR-to-X-ray SED.

The analysis of the high S/N, but rather featureless FORS spectrum showed weak \mgii{} and \civ{} in absorption, providing $\log N$(\mgii{}$)=12.96^{+0.13}_{-0.18}$. The constraint on the \mgii{} column density, together with the upper limits for the other ions and the evidence for low metallicity from the SED, suggested that the GRB absorber is most likely a $0.03\,Z_\odot <Z<1.3\,Z_\odot$ Lyman limit system, also observed along the line of sight to five other GRBs. The comparison of the \grb{} measurements with the rest-frame $EW$s of the GRB afterglow sample of \citet{Fynbo09} revealed that \grb{} is not unique, but, simply falls at the low end of the $EW$ distribution. Furthermore, we selected two sub-samples of afterglows possibly similar to \grb{}, the spectral-line poor and the low $N$(H). Most spectral-line poor afterglows show a low $A_V$ and low $N$(H), suggesting that a short distance in the host galaxy is traversed by the line of sight. In any case, a quite featureless spectrum is not as uncommon for GRB afterglows as previously claimed.

We demonstrated that the few and weak features in the spectrum of \grb{} can be the result of the particularly intense afterglow radiation ionizing a low $\log N$(\hi{})~$\sim19$ through a line of sight traversing not more than about 0.5 kpc of its faint host galaxy. We showed that few and weak spectral features are not so uncommon in GRB afterglows, including the nearby GRB\,101219B at $z=0.55$, with associated SN, for which a sub-luminous host galaxy has been claimed. Finally, the non-detection of a host galaxy at the \grb{} position indicates that an underlying galaxy, if present, must be among the faintest GRB host galaxies.

Given that about half of the GRBs in the spectral-line poor sub-sample have a detected host galaxy, it seems likely that these bursts are also located in the outskirts of a gas-rich, massive star-forming region inside its small and faint host galaxies, rather than in a halo environment.

\section*{Acknowledgments}
We thank Bradlay Cenko for sharing results prior to publication, Stephanie Courty, Jason Prochaska and Sandra Savaglio for insightful discussions, and the and the anonymous referee for a very constructive report, which significantly improved the paper. ADC acknowledges the support of the University of Iceland Research Fund and the European Commission under a Marie Curie Host Fellowship for Early Stage Researchers Training / Centre of Excellence for Space, Planetary and Astrophysics Research Training and Networking (SPARTAN, No. MEST-CT-2004-007512) hosted by the University of Leicester. RLCS is supported by a Royal Society Fellowship. KW and RLCS acknowledge support from the STFC. The financial support of the British Council and Platform Beta Techniek through the Partnership Programme in Science (PPS WS 005) is gratefully acknowledged. AJvdH was supported by NASA grant NNH07ZDA001-GLAST. PJ acknowledges support by a Marie Curie European Re-integration Grant within the 7th European Community Framework Program and a Grant of Excellence from the Icelandic Research Fund. The Dark Cosmology Centre is funded by the Danish National Research Foundation. 

\bibliographystyle{mn2e}

\bibliography{biblio}

\bsp

\appendix
\section{}
\label{appendix}

\begin{table*}
\centering
\caption{\textit{Swift} UVOT photometry of the afterglow, corrected for Galactic extinction.} 
\begin{tabular}{ c c c c }
\hline \hline
\rule[-0.2cm]{0mm}{0.6cm}
          Time     &     Brightness    &   Error   & Filter\\    
\rule[-0.2cm]{0mm}{0.6cm}
         (days)    &        (mag)      &    (mag)  & \\
\hline
\rule[-0.0cm]{0mm}{0.4cm}
       0.545   &    18.30  &    0.10 &   $v$      \\
       0.555   &    18.72  &    0.07 &   $b$      \\
       0.611   &    18.61  &    0.12 &   $v$      \\
       0.622   &    18.89  &    0.07 &   $b$      \\
       0.679   &    18.46  &    0.11 &   $v$      \\
       0.689   &    18.91  &    0.08 &   $b$      \\
       0.736   &    18.87  &    0.16 &   $v$      \\
       1.336   &    18.70  &    0.15 &   $uvw1$   \\
       1.340   &    18.69  &    0.17 &   $u$      \\
       1.349   &    18.90  &    0.11 &   $uvw2$   \\
       1.360   &    18.37  &    0.13 &   $uvm2$   \\
       1.404   &    19.00  &    0.18 &   $uvw1$   \\
       1.407   &    18.54  &    0.16 &   $u$      \\
       1.416   &    18.77  &    0.10 &   $uvw2$   \\
       1.427   &    18.76  &    0.16 &   $uvm2$   \\
       1.473   &    18.76  &    0.16 &   $uvw1$   \\
       1.476   &    18.73  &    0.19 &   $u$      \\
       1.484   &    19.09  &    0.13 &   $uvw2$   \\
       1.492   &    19.15  &    0.32 &   $uvm2$   \\
       1.743   &    19.03  &    0.25 &   $uvw1$   \\
       1.749   &    19.29  &    0.20 &   $uvw2$   \\
       1.755   &    19.08  &    0.28 &   $uvm2$   \\
       1.810   &    19.21  &    0.30 &   $uvw1$   \\
       1.816   &    19.12  &    0.18 &   $uvw2$   \\
       1.822   &    18.83  &    0.24 &   $uvm2$   \\
       1.883   &    19.71  &    0.28 &   $uvw2$   \\
       1.889   &    19.21  &    0.31 &   $uvm2$   \\
       1.944   &    18.71  &    0.21 &   $uvw1$   \\
       1.946   &    18.80  &    0.27 &   $u$      \\
       1.950   &    19.17  &    0.19 &   $uvw2$   \\
       1.956   &    19.12  &    0.30 &   $uvm2$   \\
       2.017   &    20.06  &    0.35 &   $uvw2$   \\
       2.084   &    19.42  &    0.22 &   $uvw2$   \\
       2.149   &    19.61  &    0.24 &   $uvw2$   \\
       2.155   &    19.40  &    0.35 &   $uvm2$   \\
       2.285   &    19.88  &    0.31 &   $uvw2$   \\
       2.348   &    19.16  &    0.36 &   $u$      \\
       2.421   &    19.53  &    0.25 &   $uvw2$   \\
       2.622   &    20.20  &    0.36 &   $uvw2$   \\
       2.757   &    20.26  &    0.36 &   $uvw2$   \\
       3.488   &    19.56  &    0.34 &   $uvw1$   \\
   \rule[-0.2cm]{0mm}{0.4cm}
       3.502   &    19.53  &    0.34 &   $uvm2$   \\

\hline \hline
\end{tabular}
\label{tab_uvot}
\end{table*}

\begin{figure}
\centering
\includegraphics[width=88mm,angle=0]{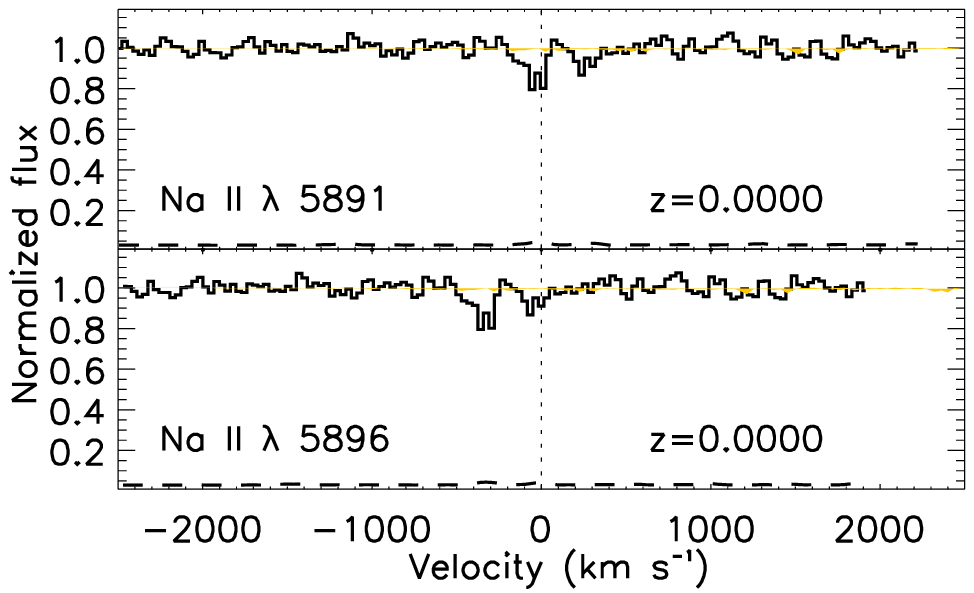}
\caption{Galactic absorption lines detected in the FORS spectra. The error spectrum is displayed at the bottom of each panel (dashed line). The telluric features are highlighted.}
\label{fig Gal}
\end{figure}

\begin{table*}
\caption{NIR-to-X-ray SED modelling with a single power law (PL), a broken power law (BPL) and a tied broken power law (TBPL), where the spectral slopes are tied to differ by $\Delta\beta\equiv0.5$, as expected for a synchrotron cooling break \citep{Sari98}. Individual epochs (1.0, 1.4, 1.9, 2.5 and 2.9 days after the trigger) are modelled separately. Quoted errors refer to a 90\% confidence level.}
\begin{tabular}{ l | c l c c l c c c }
\hline \hline
\rule[-0.2cm]{0mm}{0.6cm}
Epoch &  Ext. &  Model &  $\chi^2_\nu$ [dof]  & $E(B-V)$ & $N$(H)$_{\rm X}$ & $\beta_1$ & $\beta_2$ & $E_{\rm bk}$  \\
\rule[-0.2cm]{0mm}{0.6cm}
      &  type &        &                &  (mag) & ($10^{22}$ cm$^{-2}$)& &                &   (keV)      \\ 
\hline
\rule[-0.0cm]{0mm}{0.4cm}

    &           & PL         & 10.71 [49]  & 0.052$^{+0.009}_{-0.009}$  & $<$7.8e-07 & 0.981$^{+0.008}_{-0.007}$   &       --        &      --       \\
    &    SMC    & BPL     & 10.01 [47]    & 0.055$^{+0.008}_{-0.007}$  & $<$7.8e-07 & 0.79$^{+0.04}_{-0.05}$   & 1.03$^{+0.01}_{-0.01}$  &  0.006$^{+0.001}_{-0.001}$ \\
    &           & TIED BPL   & 10.64 [48]    & 0.069$^{+0.008}_{-0.008}$  & $<$3.5e-06 &      $\beta_2 - 0.5$      & 1.05$^{+0.01}_{-0.01}$  & 0.0035$^{+0.0005}_{-0.0004}$\\

    &           & PL         & 10.29 [49]    & 0.055$^{+0.008}_{-0.008}$  & $<$7.8e-07 & 0.986$^{+0.008}_{-0.007}$   &       --        &      --        \\
I   &    LMC    & BPL     &  9.61 [47]    & 0.059$^{+0.008}_{-0.007}$  & $<$7.8e-07 & 0.79$^{+0.05}_{-0.05}$    & 1.04$^{+0.01}_{-0.01}$  &  0.006$^{+0.002}_{-0.001}$   \\
    &           & TIED BPL   & 10.10 [48]    & 0.074$^{+0.009}_{-0.008}$  & $<$3.2e-06 &       $\beta_2 - 0.5$        & 1.05$^{+0.01}_{-0.01}$  & 0.0031$^{+0.0005}_{-0.0006}$ \\

    &           & PL         & 10.16 [49]    & 0.062$^{+0.009}_{-0.009}$  & $<$7.8e-07 & 0.989$^{+0.007}_{-0.008}$   &       --        &      --         \\
    &    MW     & BPL     &  9.65 [47]    & 0.066$^{+0.008}_{-0.008}$  & $<$7.8e-07 & 0.82$^{+0.04}_{-0.05}$   & 1.03$^{+0.01}_{-0.01}$  & 0.006$^{+0.002}_{-0.001}$   \\
\rule[-0.2cm]{0mm}{0.4cm}
    &           & TIED BPL   & 10.07 [48]    & 0.087$^{+0.008}_{-0.008}$  & $<$7.8e-07 &       $\beta_2 - 0.5$        & 1.038$^{+0.008}_{-0.009}$  & 0.0024$^{+0.0002}_{-0.0002}$ \\

\hline
\rule[-0.0cm]{0mm}{0.4cm}
    &           & PL         &  7.62 [48]    & 0.061$^{+0.007}_{-0.007}$  & $<$9.3e-07 & 1.046$^{+0.008}_{-0.008}$   &       --        &      --         \\
    &    SMC    & BPL     &  6.46 [46]    & 0.063$^{+0.007}_{-0.007}$  & $<$9.3e-07 & 1.054$^{+0.008}_{-0.009}$   & -0.3$^{+0.9}_{-0.4}$  & 3.8$^{+0.4}_{-1.7}$ \\
    &           & TIED BPL   &  7.79 [47]    & 0.061$^{+0.004}_{-0.004}$  & $<$9.3e-07 &     $\beta_2 - 0.5$          & 1.546$^{+0.004}_{-0.004}$  & 11.0$^{+1.0}_{-4.6}$ \\

    &           & PL         & 40.71 [48]    & 0.04$^{+0.01}_{-0.01}$  & 0.11$^{+0.05}_{-0.05}$ & 1.01$^{+0.01}_{-0.01}$   &       --        &      --         \\
II  &    LMC    & BPL     &  6.15 [46]    & 0.071$^{+0.007}_{-0.007}$  & $<$9.3e-07 &1.064$^{+0.009}_{-0.008}$  & 0.5$^{+0.2}_{-1.2}$  & 2.3$^{+0.4}_{-0.4}$\\
    &           & TIED BPL   &  7.53 [47]    & 0.066$^{+0.004}_{-0.004}$  & $<$9.3e-07 &     $\beta_2 - 0.5$          & 1.551$^{+0.004}_{-0.008}$  & 9.7$^{+2.2}_{-3.4}$ \\

    &           & PL         &  7.21 [48]    & 0.091$^{+0.009}_{-0.009}$  & $<$9.3e-07 & 1.059$^{+0.008}_{-0.009}$   &       --        &      --        \\
    &    MW     & BPL     &  5.87 [46]    & 0.098$^{+0.009}_{-0.009}$  & $<$9.3e-07 & 1.073$^{+0.009}_{-0.009}$  & 0.49$^{+0.15}_{-1.02}$  & 2.3$^{+0.4}_{-0.4}$\\
\rule[-0.2cm]{0mm}{0.4cm}
    &           & TIED BPL   &  7.36 [47]    & 0.091$^{+0.005}_{-0.005}$  & $<$9.3e-07 &     $\beta_2 - 0.5$          & 1.559$^{+0.004}_{-0.005}$  & 10.4$^{+1.6}_{-4.0}$ \\

\hline
\rule[-0.0cm]{0mm}{0.4cm}
    &           & PL         &  7.62 [46]    & $<$0.005 & 0.09$^{+0.07}_{-0.06}$  & 1.074$^{+0.009}_{-0.004}$   &       --        &      --        \\
    &    SMC    & BPL     &  7.86 [44]    & $<$0.028 & 0.13$^{+0.14}_{-0.09}$  & 0.9$^{+0.1}_{-0.1}$   & 1.12$^{+0.16}_{-0.04}$  & 0.006$^{+0.131}_{-0.002}$ \\
    &           & TIED BPL   &  7.84 [45]    & 0.027$^{+0.013}_{-0.013}$ & 0.26$^{+0.13}_{-0.12}$  &     $\beta_2 - 0.5$          & 1.25$^{+0.08}_{-0.08}$  & 0.015$^{+0.038}_{-0.010}$ \\
 
    &           & PL         &  7.62 [46]    & $<$0.007 & 0.09$^{+0.07}_{-0.06}$  & 1.074$^{+0.010}_{-0.004}$   &      --         &     --      \\
III &    LMC    & BPL     &  7.82 [44]    & $<$0.011 & 0.08$^{+0.08}_{-0.06}$  & -0.183$^{+0.070}_{-0.003}$   & 1.076$^{+0.013}_{-0.008}$ &0.00084$^{+0.00050}_{-0.00002}$ \\
    &           & TIED BPL   &  3.62 [45]    & 0.03$^{+0.01}_{-0.01}$ & $<$4.9e-06  &    $\beta_2 - 0.5$           & 1.19$^{+0.01}_{-0.02}$   & 0.00462$^{+0.0010}_{-0.0008}$ \\
 
    &           & PL         &  7.62 [46]    & $<$0.007 & 0.09$^{+0.07}_{-0.06}$  & 1.074$^{+0.009}_{-0.004}$   &      --         &    --       \\
    &    MW     & BPL     &  7.88 [44]    & $<$0.026 & $<$0.13  & 0.98$^{+0.08}_{-0.05}$  & 1.11$^{+8.e-08}_{-0.01}$ & 0.013$^{+0.005}_{-0.009}$\\
\rule[-0.2cm]{0mm}{0.4cm}
    &           & TIED BPL   &  7.97 [45]    & 0.026$^{+0.019}_{-0.020}$ & 0.31$^{+0.13}_{-0.12}$  &    $\beta_2 - 0.5$        & 1.30$^{+0.08}_{-0.08}$  & 0.03$^{+0.08}_{-0.02}$ \\

\hline
\rule[-0.0cm]{0mm}{0.4cm}
    &           & PL         &  8.95 [48]    & $<$0.003 &0.11$^{+0.09}_{-0.08}$  & 1.083$^{+0.}_{-0.003}$  &     --         &     --     \\
    &    SMC    & BPL     &  8.55 [46]   & $<$0.023 &0.20$^{+0.17}_{-0.12}$  & 0.48$^{+0.13}_{-0.14}$   & 1.19$^{+0.13}_{-0.04}$ & 0.006$^{+0.009}_{-0.001}$ \\
    &           & TIED BPL   &  2.05 [47]   & 0.019$^{+0.012}_{-0.012}$ & $<$2.4e-06  &    $\beta_2 - 0.5$            & 1.188$^{+0.020}_{-0.017}$ &  0.005$^{+0.002}_{-0.001}$\\
  
    &           & PL         &  8.95 [48]   & $<$0.003 & 0.11$^{+0.09}_{-0.08}$ & 1.083$^{+0.}_{-0.003}$  &     --         &     --      \\
IV  &    LMC    & BPL     &  8.48 [46]   & 0.025$^{+0.015}_{-0.013}$ & 0.20$^{+0.16}_{-0.11}$ & 0.65$^{+0.10}_{-0.12}$  & 1.20$^{+0.12}_{-0.03}$ & 0.006$^{+0.011}_{-0.001}$\\
    &           & TIED BPL   &  1.99 [47]   & 0.020$^{+0.011}_{-0.010}$ & $<$2.9e-06  &    $\beta_2 - 0.5$            & 1.189$^{+0.018}_{-0.016}$ & 0.005$^{+0.002}_{-0.001}$  \\
 
    &           & PL         &  8.95 [48]   & $<$0.003 &0.11$^{+0.09}_{-0.08}$  & 1.083$^{+0.}_{-0.003}$  &     --         &     --       \\
    &    MW     & BPL     &  8.49 [46]   & 0.023$^{+0.013}_{-0.012}$ & 0.20$^{+0.16}_{-0.11}$ & 0.67$^{+0.10}_{-0.11}$  & 1.19$^{+0.12}_{-0.04}$ &0.006$^{+0.012}_{-0.002}$ \\
\rule[-0.2cm]{0mm}{0.4cm}
    &           & TIED BPL   &  3.69 [47]   & $<$0.009 & $<$2.2e-06 &     $\beta_2 - 0.5$           & 1.590$^{+0}_{-0.006}$  & 8.1$^{+3.9}_{-5.5}$\\

\hline
\rule[-0.0cm]{0mm}{0.4cm}
    &           & PL         &  7.87 [48]   &$<$0.002 & 0.10$^{+0.10}_{-0.09}$ & 1.062$^{+0.}_{-0.003}$  &     --         &      --     \\
    &    SMC    & BPL     &  6.73 [46]   &0.023$^{+0.016}_{-0.015}$ & 0.20$^{+0.17}_{-0.11}$ & 0.67$^{+0.11}_{-0.11}$  & 1.19$^{+0.12}_{-0.04}$ & 0.006$^{+0.012}_{-0.002}$ \\
    &           & TIED BPL   &  6.69 [47]   &$<$0.007 & 0.16$^{+0.10}_{-0.09}$ &     $\beta_2 - 0.5$           & 1.145$^{+0.}_{-0.003}$ & 0.0057$^{+0.0006}_{-0.0007}$ \\
  
    &           & PL         &  7.87 [48]   & $<$0.001& 0.10$^{+0.10}_{-0.09}$ & 1.062$^{+0}_{-0.003}$  &     --         &      --      \\
V   &    LMC    & BPL     &  6.73 [46]   &$<$0.0157 & 0.20$^{+0.17}_{-0.12}$ & 0.48$^{+0.13}_{-0.14}$  & 1.19$^{+0.13}_{-0.04}$ & 0.006$^{+0.009}_{-0.001}$\\
    &           & TIED BPL   &  6.69 [47]   &$<$0.006 & 0.16$^{+0.10}_{-0.09}$ &     $\beta_2 - 0.5$           & 1.145$^{+0.}_{-0.003}$ & 0.0057$^{+0.0006}_{-0.0008}$ \\
 
    &           & PL         &  7.87 [48]   &$<$0.001 &0.10$^{+0.10}_{-0.09}$  & 1.062$^{+0.}_{-0.003}$  &      --        &      --       \\
    &    MW     & BPL     &  6.73 [46]   &$<$0.014 & 0.20$^{+0.18}_{-0.12}$ & 0.49$^{+0.12}_{-0.13}$  &1.18$^{+0.13}_{-0.04}$  & 0.006$^{+0.008}_{-0.001}$ \\
\rule[-0.2cm]{0mm}{0.4cm}
    &           & TIED BPL   &  6.69 [47]   & $<$0.005& 0.16$^{+0.10}_{-0.09}$ &     $\beta_2 - 0.5$           & 1.145$^{+0.}_{-0.003}$ & 0.0057$^{+0.0007}_{-0.0007}$ \\

\hline \hline
\end{tabular}
\label{isis_table}
\end{table*}

\label{lastpage}
\end{document}